\documentclass[pra,aps,twocolumn,footinbib,floatfix,10pt,longbibliography,superscriptaddress]{revtex4-1}

\usepackage[english]{babel}
\usepackage{breakcites}

\usepackage[utf8]{inputenc}
\usepackage[T1]{fontenc}
\usepackage[bookmarks=false]{hyperref}
\usepackage{graphicx}
\usepackage{braket}
\usepackage{color}
\usepackage{epstopdf}
\usepackage{mathtools}
\usepackage{amsmath}
\usepackage{amssymb}
\usepackage[bookmarks=false]{hyperref}
\usepackage{float}
\usepackage{gensymb}
\usepackage[colorinlistoftodos,prependcaption]{todonotes}
\usepackage{xargs}  
\newcommandx{\unsure}[2][1=]{\todo[linecolor=red,backgroundcolor=red!25,bordercolor=red,#1]{#2}}
\newcommandx{\change}[2][1=]{\todo[linecolor=blue,backgroundcolor=blue!25,bordercolor=blue,#1]{#2}}
\newcommandx{\info}[2][1=]{\todo[linecolor=OliveGreen,backgroundcolor=OliveGreen!25,bordercolor=OliveGreen,#1]{#2}}
\newcommandx{\improvement}[2][1=]{\todo[linecolor=Plum,backgroundcolor=Plum!25,bordercolor=Plum,#1]{#2}}
\newcommandx{\thiswillnotshow}[2][1=]{\todo[disable,#1]{#2}}
\newcommandx{\greencom}[2][1=]
{\todo[inline, color=green!40,#1]{#2}}
\newcommandx{\bluecom}[2][1=]
{\todo[inline, color=blue!40,#1]{#2}}

\usepackage{letltxmacro}

\LetLtxMacro{\ORIGselectlanguage}{\selectlanguage}
\makeatletter
\DeclareRobustCommand{\selectlanguage}[1]{%
  \@ifundefined{alias@\string#1}
    {\ORIGselectlanguage{#1}}
    {\begingroup\edef\x{\endgroup
       \noexpand\ORIGselectlanguage{\@nameuse{alias@#1}}}\x}%
}
\newcommand{\definelanguagealias}[2]{%
  \@namedef{alias@#1}{#2}%
}
\makeatother

\definelanguagealias{en}{english}
\definelanguagealias{EN}{english}

\begin{document}
\emergencystretch 3em
\raggedbottom
\abovedisplayskip=8pt
\abovedisplayshortskip=0pt
\belowdisplayskip=8pt
\belowdisplayshortskip=8pt
\newcommand{\vc}[1]{{\boldsymbol{\mathrm{#1}}}}
\newcommand{\comment}[1]{}
\newcommand{\remove}[1]{}
\newcommand{\quot}[1]{\textquotedblleft{}#1\textquotedblright}
\newcommand{\un}{\mathrm}
\newcommand{\B}{\langle B \rangle}
\newcommand{\blue}[1]{{\color{blue}#1}}
\newcommand{\red}[1]{{\color{red}#1}}
\newcommand{\orange}[1]{{\color{orange}#1}}
\makeatletter
\newcommand{\vast}{\bBigg@{4}}
\newcommand{\Vast}{\bBigg@{5}}
\newcommand{\G}{\mathbf{G}}
\makeatother
\title{Efficient Pulse-Excitation Techniques for Single Photon Sources from Quantum Dots in Cavities}
\author{Chris Gustin} 
\email{c.gustin@queensu.ca}
\affiliation{Department of Physics, Engineering Physics, and Astronomy, Queen's University, Kingston, Ontario K7L 3N6, Canada}
\author{Stephen Hughes} 
\affiliation{Department of Physics, Engineering Physics, and Astronomy, Queen's University, Kingston, Ontario K7L 3N6, Canada}

\begin{abstract}
We present rigorous and intuitive master equation models to study on-demand single photon sources
from pulse-excited quantum dots coupled to cavities. We consider three methods of source excitation: resonant pi-pulse, off-resonant phonon-assisted inversion, and two-photon excitation of a biexciton-exciton cascade, and investigate the effect of the pulse excitation process on the quantum indistinguishability, efficiency, and purity of emitted photons. By explicitly modelling the time-dependent pulsed excitation process in a manner which captures non-Markovian effects associated with coupling to photon and phonon reservoirs, we find that photons of near-unity indistinguishability can be emitted with over $90\%$ efficiency for all these schemes, with the off-resonant schemes not necessarily requiring polarization filtering due to the frequency separation of the excitation pulse, and allowing for very high single photon purities. Furthermore, the off-resonant methods are shown to be robust over certain parameter regimes, with less stringent requirements on the excitation pulse duration in particular. We also derive a semi-analytical simplification of our master equation for the off-resonant drive, which gives insight into the important role that exciton-phonon decoupling for a strong drive plays in the off-resonant phonon-assisted inversion process. 
\end{abstract}
\date{\today}

\maketitle

\section{Introduction}
Recently, the field of semiconductor quantum dot (QD)-cavity single photons sources (SPSs) has received much attention, with theoretical analyses spanning a broad range of topics, and numerous high quality experimental sources being realized in the past few years.
In terms of understanding how to improve QD SPSs~\cite{kiraz04,michler00}, much effort has been focused on the impact of electron-phonon scattering on the SPS efficacy~\cite{ilessmith17,gustin17,grange17,ross16,tighineanu18,denning18,gerhardt18}, which degrades the SPS figures-of-merit via incoherent phonon relaxation-assisted photon emission, as well as the effects of the conventional resonant excitation process~\cite{gustin_pulsed_2018}---the most notable of which being the potential for multi-photon emission events from a single pulsed excitation~\cite{fischer17,2fischer17}. The increased theoretical understanding of how to realize high SPS figure-of-merit as well as experimental advances in charge noise reduction~\cite{reed16,kuhlmann13} have led to numerous photon sources in recent years with high efficiency, purity (lack of multiphoton events), and quantum indistinguishability of emitted photons~\cite{somaschi16,senellart17,ding16,hanschke18,schweickert18,dusanowski19,liu18}. Waveguide-based photon sources have also seen extensive development~\cite{dusanowski19,daveau17,zadeh16,laucht12}, and integrated cavity-waveguide systems have the potential to harness the advantages of cavity coupling~\cite{yao10,liu18}.



A commonly-held belief is that resonantly pumped sources are preferable to off-resonantly excited ones (in which the source is populated by decay from higher lying energy levels), due to the so-called timing jitter that arises from radiative emission from higher-lying states to the exciton of interest, which degrades the phase coherence of the emitted single photons, leading to a SPS which produces photons of poor indistinguishability~\cite{senellart17,kiraz04}. Furthermore, resonant pulsed excitation leads to deterministic and \emph{on-demand} production of single photons, which is critical for many potential applications. However, the fundamental requirement is not strictly resonant excitation, but rather that the (on-demand) excitation process creates an excited state in the QD in a deterministic and \emph{coherent} manner---indeed, other (potentially) effective methods of coherent single-photon source generation include off-resonant phonon-assisted exciton inversion~\cite{quilter15,ross16,reindl19}, modified STIRAP~\cite{gustin17}, two-photon excitation of a biexciton state~\cite{hanschke18}, and adiabatic rapid passage via a chirped pulse~\cite{wei14,reuble14,reiter14}. As the QD-cavity SPS rapidly advances towards a scalable, high-fidelity implementation, there is a clear need to understand the role of the excitation process on the quantum dynamics (particularly with respect to the phonon and photon reservoir couplings) and SPS figures-of-merit for these different excitation schemes. While the resonantly inverted QD-cavity SPS has been studied theoretically to these ends~\cite{gustin_pulsed_2018}, analyses of other excitation methods have often been restricted to studying population dynamics via one-time correlation functions of the system, which are sufficient for determining the SPS efficiency (via emitted photon numbers), but insufficient for calculation of the indistinguishability or purity of the emitted photons, which typically require two-time correlation functions. Furthermore, the need for polarization filtering for a resonant pulse (which reduces the effective efficiency by at least $50\%$), as well as the requirement of very short pulses to suppress multiphoton emission, naturally leads one to consider alternative methods of SPS generation. 

In this work, we present rigorous time-dependent master equation (ME) models of the exciton-phonon and exciton-cavity interactions for a driven QD, valid for the short and high drive strength pulses which are often required for effective SPS generation. We focus on three methods of SPS excitation---resonant pulse, off-resonant acoustic phonon-assisted pulse, and biexciton excitation via a two-photon resonance process---and compare the impact of the excitation process on the SPS figures-of-merit (efficiency, indistinguishability, and single photon purity) for these methods. We extend the work of Ref.~\cite{ross16}, which partly studied the figures-of-merit for a SPS excited via off-resonant phonon-assisted excitation with a phonon ME, but did not correctly include the effect of multi-photon generation, and used a ``bad-cavity" approximation with respect to the cavity dynamics, which are now known to fail to correctly produce the dynamical decoupling effects associated with a short pulse~\cite{gustin_pulsed_2018}.  
In addition, we derive a semi-analytical simplification of the ME for the off-resonantly driven exciton, which provides explicit expressions for the relevant phonon scattering rates that are responsible for the exciton inversion, as well as drive-dependent pure dephasing. The layout of the rest of the paper is as follows: in Section~\ref{models}, we derive the time-convolutionless weak phonon coupling MEs used in the analysis of our QD-cavity setups, in Section~\ref{results} we present the results of our numerical simulations on the figures-of-merit for these excitation methods, as well as some insight into the underlying physics, and in Section~\ref{conclusions} we conclude. We also include two appendices, which contain a comparison of our weak phonon coupling ME with a polaron transform ME approach, and the full analytical simplification of the single-exciton ME, respectively.

\section{Quantum Dot - Cavity Models}\label{models}

\begin{figure}[b]
\centering
\includegraphics[width=1\linewidth]{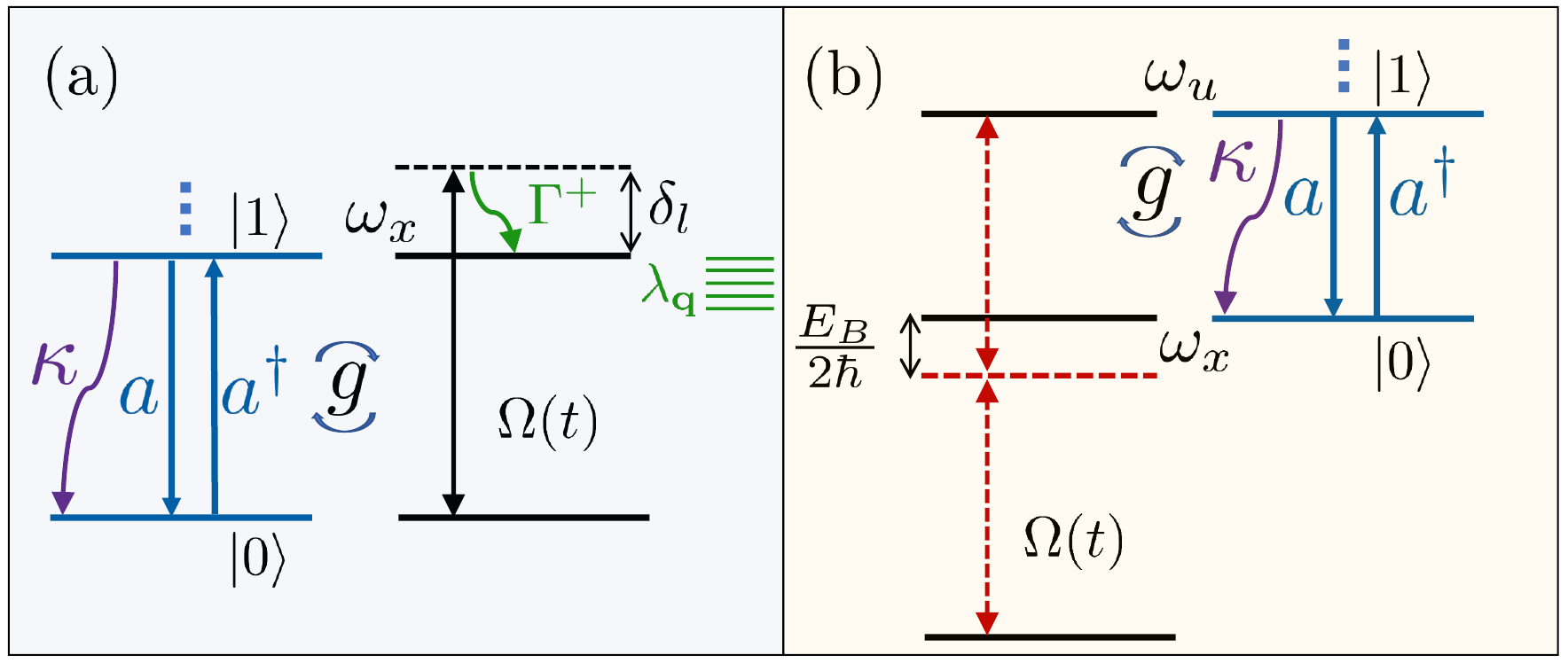}
\caption{\small  (a) Energy-level schematic of QD-cavity system for a single exciton coupled to a cavity mode with effective laser detuning $\delta_l$. For $\delta_l=0$, this corresponds to resonant excitation, and for $\delta_l >0$, this corresponds to off-resonant phonon-assisted excitation, where $\Gamma^+$ is the dominant scattering rate corresponding to this process. (b) Schematic of QD-cavity system for two-photon excitation of the biexciton, where the cavity is resonant to the biexciton-exciton transition (phonon states not shown for simplicity). Not shown here is another orthogonally polarized exciton state, $\ket{y}$, which is populated only via spontaneous emission from the biexciton. 
} 
\label{fig_s} 
\end{figure}

In this section, we describe the relevant quantum optics models of our QD-cavity system, and derive weak phonon coupling MEs to treat the electron-phonon interaction. To describe the case of a resonantly or off-resonantly (i.e., phonon-assisted inversion) pumped QD, we consider a single two-level system with ground $\ket{g}$ and exciton $\ket{x}$ states resonantly coupled to a quantized cavity mode with bosonic creation and destruction operators $a^{\dagger}$, $a$. We also study biexciton preparation via two-photon excitation (TPE) by considering a four-level biexciton cascade setup (two linearly polarized excitons and a two exciton, or biexciton, state), where the biexciton to exciton transition is coupled to the cavity. Figure~\ref{fig_s} gives a schematic of these two scenarios. We then derive a time-convolutionless ME for this pulsed QD-cavity system, by treating the electron-phonon interaction perturbatively, and performing a Born-Markov approximation. This results in a  \emph{weak phonon coupling ME}. Such an approach misses certain non-Markovian features of the electron-phonon scattering, including the broad phonon sidebands~\cite{roy11,roy12}. However, since we are dealing with photons emitted from the cavity mode, the cavity acts as a Lorentzian filter around the cavity mode resonance, and we only need the Markovian phonon dynamics to compute the output observables. While the weak coupling approach breaks down at elevated temperatures, we restrict our analysis to cryogenic temperatures $T = 4 \ \rm{K}$, where the weak phonon coupling ME is appropriate for realistic QDs (as we show directly via a comparison with a polaron transform ME approach in Appendix~\ref{AppendixA}). The advantage of the weak phonon coupling ME is that it remains valid as the pump pulse intensity increases, and is thus capable of predicting the exciton-phonon decoupling that occurs at strong enough drive strengths, whereas a polaron ME approach fails~\cite{mccutcheon11,mccutcheon10}. This will prove to be relevant in some of the regimes studied below.

\subsection{Pulse excitation of a single exciton-cavity system}\label{modelsA}
In this subsection, we consider the model used to study resonant pi-pulse SPSs, as well as the off-resonant phonon-assisted inversion SPS, by considering the single exciton system shown in Figure~\ref{fig_s}a, driven by a pulse with time-dependent amplitude $\Omega(t) = \Omega_0 e^{-\left(\frac{t-3\tau_p}{\tau_p}\right)^2}$, where $\tau_p$ is the pulse width. Our approximation of neglecting higher lying states is appropriate for $\frac{\tau_p E_B}{4\hbar} \gg 1$ (see Figure~\ref{fig_s}b) for neutral QDs (such that the frequency-domain pulse amplitude at  $\omega_L - \frac{E_B}{2\hbar}$ 
is much less than its maximum amplitude), or charged QDs (trion states)~\cite{trivedi19}.


Our first system of interest consists of a two-level neutral QD system with ground and exciton states, described with Pauli pseudospin  operators $\sigma^{\pm}$ ($\sigma_x = \sigma^+ + \sigma^-$, $\sigma_y = i(\sigma^- - \sigma^+)$), a cavity mode, and a reservoir of bosonic phonon modes with operators $b_{\mathbf{q}}$, $b^{\dagger}_{\mathbf{q}}$ for wavevector $\mathbf{q}$. Neglecting interactions with the photon bath reservoirs for now, we begin with the total Hamiltonian  $H = H_S + H_B + H_I$ for a two-level system (below we also model the biexciton cascade setup) driven by a laser with time-dependent Rabi drive $\Omega(t)$ in frame rotating at the laser frequency $\omega_L$, where
\begin{align}\label{Hs}
H_S \!= \!  \hbar\Delta_x\sigma^+ \sigma^- \!+\!  \hbar\Delta_ca^\dagger a \!+\! \frac{\hbar\Omega(t)}{2}\sigma_x \!+\! \hbar g(\sigma^{+}a \!+\! \sigma^{-}a^{\dagger}),
\end{align}
\begin{equation}\label{Hb}
H_B=\sum\limits_{\mathbf{q}}\hbar\omega_{\mathbf{q}}b_{\mathbf{q}}^{\dagger}b_{\mathbf{q}},
\end{equation}
and
\begin{equation}\label{Hi}
     H_I=\sigma^+ \sigma^-\sum\limits_{\mathbf{q}}\hbar\lambda_{\mathbf{q}}(b_{\mathbf{q}}^{\dagger}+b_{\mathbf{q}})
\end{equation}
are the system, phonon bath, and interaction Hamiltonians, respectively.
Here, the laser detuning from the exciton frequency $\omega_x$ is $\Delta_x =  \omega_x - \omega_L$, and the cavity detuning $\Delta_c = \omega_c - \omega_L$. We assume the cavity is resonant with the exciton in this subsection. In the continuum limit of phonon modes, we can characterize the electron--acoustic-phonon interaction with the phonon spectral function appropriate for exciton-phonon interactions via a deformation potential in QDs: $J_p(\omega) = \sum_{\mathbf{q}}\lambda_{\mathbf{q}}^2\delta(\omega-\omega_{\mathbf{q}}) \rightarrow J_p(\omega) = \alpha \omega^3 \text{exp}\big[\!-\!\frac{\omega^2}{2\omega_b^2}\big]$, where $\alpha$ characterizes the exciton-phonon coupling strength, and $\omega_b$ is a frequency cutoff which depends on the size of the QD~\cite{nazir16}. The weak phonon coupling ME, in the limit of no excitation pulse or cavity, gives a polaron shift of the exciton frequency, $\omega_x \rightarrow \omega_x - \Delta_P$, where $\Delta_p = \int_0^{\infty}d\omega J_p(\omega)/\omega = \alpha \omega_b^3 \sqrt{\pi/2}$. Thus, the resonance condition between the cavity and exciton is $\Delta_x = \Delta_p - \delta_l$ and $\Delta_c = -\delta_l$, where the effective laser detuning (magnitude) becomes $\delta_l$. From this point, we can derive a weak phonon coupling ME.

\subsection{Weak Phonon Coupling}\label{modelsB}
At low temperatures and for $\alpha \omega_b^2 \ll 1$, we can treat the interaction term $H_I$ perturbatively by performing a 2nd-order Born-Markov approximation, tracing over the phonon reservoir~\cite{mccutcheon10}. This procedure generates the \emph{weak phonon coupling ME}:
\begin{equation}\label{eq:wpme}
    \frac{\rm{d}\rho}{\rm{dt}} = -\frac{i}{\hbar}[H_S,\rho] + \frac{1}{2}\sum_{\mu}\mathcal{L}[A_\mu]\rho + \mathbb{L_{\rm{w}}}\rho,
    \end{equation}
    where we have added Lindblad terms, $\mathcal{L}[A]\rho=2A\rho A^\dagger - A^\dagger A \rho - \rho A^\dagger A$, corresponding to radiative decay from the cavity mode with rate $\kappa$ ($\mathcal{L}[\sqrt{\kappa}a]\rho$), and decay from the exciton into other background modes with rate $\gamma$ ($\mathcal{L}[\sqrt{\gamma}\sigma^-]$).
    The weak phonon coupling term corresponding to exciton-phonon scattering is 
    \begin{equation}
        \mathbb{L_{\rm{w}}}\rho = \int_0^{\infty}d\tau \Gamma_{\rm{w}}(\tau)\big(\tilde{N}(t-\tau,t)\rho N - N \tilde{N}(t-\tau,t)\rho\big) + \text{H.c.},
        \end{equation}
        where $N = \sigma^+\sigma^-$, and
        \begin{equation}
            \tilde{N}(t-\tau,t) = U^{\dagger}(t-\tau,t)N U(t-\tau,t),
        \end{equation}
        and the operator $U(t,t_0)$ evolves the system via the system Hamiltonian $H_S(t)$ from time $t_0$ to $t$. In general, calculation of this operator is non-trivial, but here we can make an additional Markov approximation with respect to the \emph{time-dependent} element of the system Hamiltonian, valid for $\tau_p\omega_b \gg 1$ as shown in Ref.~\cite{gustin_pulsed_2018}; for these parameters, $\tau_p \gtrsim 2 \ \text{ps}$. Pulse widths larger than this value additionally suppress two-photon excitation of higher energy exciton states for common QD parameters (biexciton binding energy). The phonon-scattering term then simplifies to:
           \begin{equation}\label{eq:me}
        \mathbb{L_{\rm{w}}}\rho \approx \int_0^{\infty}d\tau \Gamma_{\rm{w}}(\tau)\big(\tilde{N}(-\tau)\rho N - N \tilde{N}(-\tau)\rho\big) + \text{H.c.},
        \end{equation}
        where $\tilde{N}(-\tau) = e^{-iH_S(t)\tau/\hbar}Ne^{iH_S(t)\tau/\hbar}$, and
        \begin{equation}
            \Gamma_{\rm{w}}=\! \int_0^\infty\! d\omega J_p(\omega)\left[\coth{\left(\frac{\hbar\omega}{2 k_B T}\right)}\cos{(\omega \tau)} - i\sin{(\omega \tau)}\right].
        \end{equation}
        \emph{During} excitation by a short pulse (as is the case in this work), the dynamics dictated by Equation~\eqref{eq:me} can further be simplified by neglecting the role of the cavity in the $\tilde{N}(-\tau)$ transformation, and we use this simplification in Figures~\ref{fig3} and~\ref{fig2} as well as to derive explicit scattering rates for certain phonon processes. In  Appendix \ref{AppendixB}, we evaluate the validity of this approximation and show the full ME in this simplified form.
        
        \subsection{Two-Photon Excitation via the Biexciton-Exciton Cascade}\label{modelsC}



In this subsection, we consider the excitation of a SPS consisting of a four level biexciton-exciton cascade system;
this includes the QD (ground $\ket{g}$, linearly polarized excitons $\ket{x}$ and $\ket{y}$, and the biexciton $\ket{u}$), with the biexciton-exciton transition coupled to a cavity mode, which decreases timing jitter effects via the Purcell effect. The biexciton state is excited via TPE using a pulse that is detuned from the exciton state with  half the biexciton frequency (i.e., two-photon resonant with the biexciton state, cf.~Figure~\ref{fig_s}b). In a frame rotating at the laser frequency, the system Hamiltonian for this setup is
\begin{align}
H_S =& \frac{E_B}{2}\sigma^+\sigma^- - \frac{E_B}{2}a^{\dagger}a + \frac{\hbar\Omega(t)}{2}(\sigma_x + \sigma_x^{(u)}) \nonumber \\ &+ \hbar g (a^{\dagger}\sigma^-_u + a\sigma^+_u),
\end{align}
where $\sigma^+_u = \ket{u}\bra{x}$, $\sigma^-_u = \ket{x}\bra{u}$, and $\sigma_x^{(u)} = \sigma^+_u + \sigma^-_u$.
Note that to avoid cavity coupling to the ground to exciton transition, we require $\hbar\kappa \ll E_B$. Also, to avoid direct excitation of the exciton, the pulse width is limited by $\frac{\tau_p E_B}{4\hbar} \gg 1$. As before, the bath Hamiltonian is $H_B=\sum\limits_{\mathbf{q}}\hbar\omega_{\mathbf{q}}b_{\mathbf{q}}^{\dagger}b_{\mathbf{q}}$, and the interaction Hamiltonian is now
\begin{equation}\label{Hi}
     H_I=N_{ux}\sum\limits_{\mathbf{q}}\hbar\lambda_{\mathbf{q}}(b_{\mathbf{q}}^{\dagger}+b_{\mathbf{q}}).
\end{equation}

We take the phonon coupling constant for the biexciton state to be twice the exciton phonon coupling constant~\cite{hargart16}, which gives $N_{ux} = 2\sigma^+_u\sigma^-_u + \sigma^+\sigma^-$. We subsequently derive a similar ME for this biexciton cascade setup:
\begin{equation}
    \frac{\rm{d}\rho}{\rm{dt}} = -\frac{i}{\hbar}[H_S,\rho] + \frac{1}{2}\sum_{\mu}\mathcal{L}[A_\mu]\rho + \mathbb{L_{\rm{w}}}\rho,
    \end{equation}
    with Lindblad terms corresponding to spontaneous emission into background photonic reservoirs $\sqrt{\gamma_u/2}\sigma^-_u$, $\sqrt{\gamma_u/2}\ket{y}\bra{u}$, $\sqrt{\gamma}\sigma^-$, and $\sqrt{\gamma}\ket{g}\bra{y}$. We also, again, have loss from the cavity $\sqrt{\kappa}a$. Note that we have assumed that the background radiative decay rates are independent of the polarization ($x$ or $y$) of the transition, though this is easily relaxed. The phonon scattering term $\mathbb{L_{\rm{w}}}\rho$ is the same as Equation~\eqref{eq:me}, but with $N\rightarrow N_{ux}$, and $\tilde{N}_{ux}$ calculated with the $H_S$ from this section.

\subsection{Single Photon Source Figures-of-Merit}\label{modelsD}
To quantify the efficiency of the SPSs studied, we use the expectation value of the emitted cavity photon number: 
\begin{equation}
    N_a = \kappa\int_0^{\infty}\langle a^{\dagger}a\rangle(t)dt,
\end{equation}
where the long time limit is after the system has decayed to steady state following a single pulse excitation. We
also quantify the indistinguishability by simulation of a Hong-Ou-Mandel two-photon interference experiment, as in Ref~\cite{hughes19}:
\begin{equation}
    \mathcal{I} =1 - D_1-D_2,
\end{equation}
where
\begin{equation}
    D_1 = \frac{\int_0^{\infty} dt \int_{0}^{\infty} d\tau \big(G_{\text{pop}}^{(2)}(t,\tau)
-|G^{(1)}(t,\tau)|^2\big)}{\int_0^{\infty} dt \int_{0}^{\infty} d\tau   \left ( 2G_{\text{pop}}^{(2)}(t,\tau) - |\langle a(t+\tau)\rangle \langle a^{\dagger}(t)\rangle|^2 \right)}
\label{eq:D1}
\end{equation}
gives the degradation of the two-photon interference due to the first-order coherence (note that this can also be affected by multi-photon emission), and
 \begin{equation}
D_2 = \frac{\int_0^T dt \int_{0}^T d\tau
 G^{(2)}(t,\tau) }{\int_0^T dt \int_{0}^{T} d\tau   \left(2G_{\text{pop}}^{(2)}(t,\tau) - |\langle a(t+\tau)\rangle \langle a^{\dagger}(t)\rangle|^2\right)}
 \label{eq:D2}
\end{equation}
is the degradation of the two-photon interference due to the second-order coherence (multi-photon states). Additionally, $G_{\text{pop}}^{(2)}(t,\tau) = \langle a^{\dagger}a\rangle(t)\langle a^{\dagger}a\rangle(t+\tau)$, and the first and second order coherences are $G^{(1)}(t,\tau)=\langle a(t+\tau)a^{\dagger}(t)\rangle$ and $G^{(2)}(t,\tau)= \langle a^{\dagger}(t)a^{\dagger}(t+\tau)a(t+\tau)a(t)\rangle$, respectively, calculated via the quantum regression theorem~\cite{carmichael}.

Another useful quantity is the Purcell factor---the enhancement of the spontaneous rate into the cavity. Although only strictly accurate in a long-time limit and with weak cavity coupling ($g \ll \kappa$), this metric is given by $F_P=\frac{4g^2}{\kappa \gamma}$ (on resonance), which is slightly reduced by phonon coupling~\cite{roychoudhury15}. High Purcell factors can, even for a short pulse, simultaneously increase SPS efficiency and indistinguishability by increasing the proportion of photons emitted into the desired cavity mode, while filtering out phonon sidebands that degrade the coherence of emitted photons~\cite{ilessmith17,grange17,gustin_pulsed_2018}.

\section{Results}\label{results}

\begin{figure}[tb]
\centering
\includegraphics[width=1\linewidth]{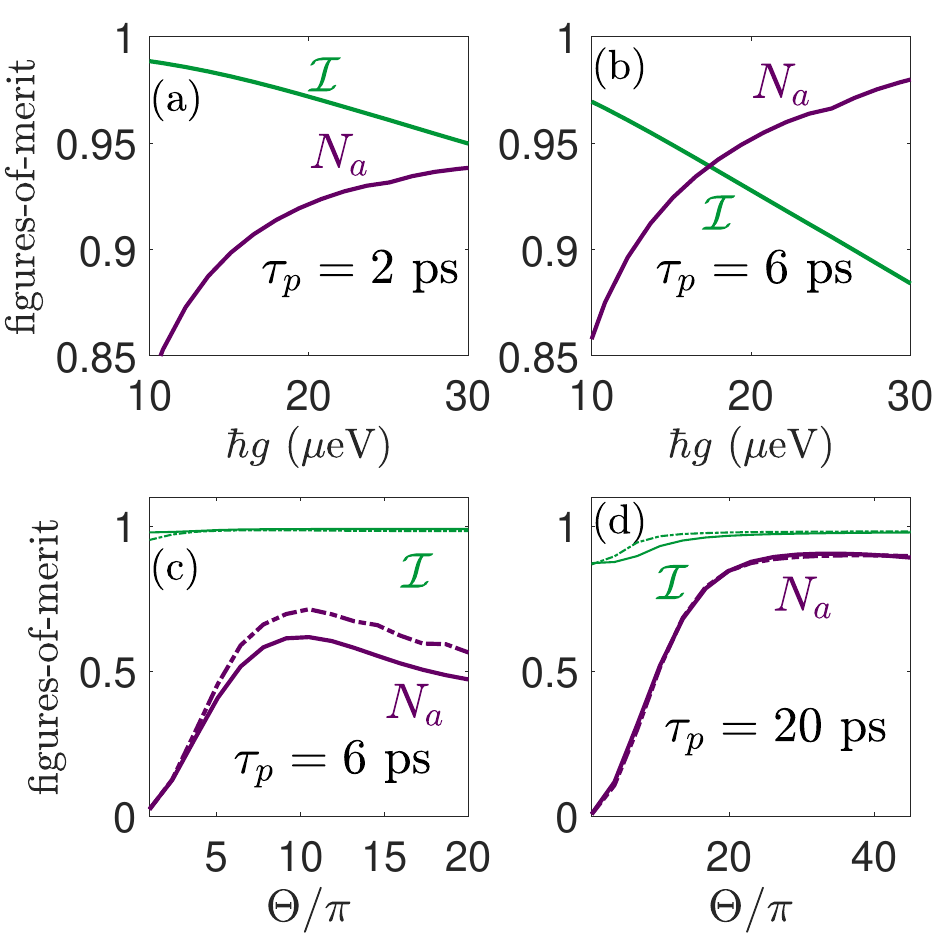}
\caption{\small Comparison of single photon indistinguishability and emitted cavity photon number for different pulse-excitation parameters under (a-b) resonant excitation and (c-d) off-resonant phonon-assisted inversion. In (c) and (d), the solid lines correspond to a detuning $\hbar\delta_l = 1 \ \text{meV}$, and the dash-dotted lines correspond to $\hbar\delta_l = 0.5 \ \text{meV}$.} 
\label{fig1} 
\end{figure}
For our numerical calculations, 
we use an optical pulse (described in Section~\ref{modelsA}) where $\Theta = \int_{-\infty}^{\infty}\Omega(t)dt$ is the pulse area, and $\tau_{\text{FWHM}} = 2\sqrt{\ln{(2)}}\tau_p$. We also use phonon parameters $\alpha = 0.03 \ \text{ps}^2$, $\hbar\omega_b = 0.9 \ \text{meV}$~\cite{quilter15}, and a temperature of $T=4 \ \text{K}$. For a resonant pulse to invert the QD, we use $\Theta = \pi$; the phonon interaction gives a coherent attenuation of the pulse strength (as well as $g$), which is captured with a polaron transform approach~\cite{mccutcheon10,roy11}, although we neglect this as it has a small effect on the dynamics for our phonon parameters and temperature ($\Omega \rightarrow 0.96 \Omega$). For all the simulations below, we let $\hbar\gamma = 1 \ \rm{\mu eV}$ ($\sim 660$ ps exciton lifetime without cavity coupling). Unless otherwise stated, our cavity parameters are $\hbar g = 20 \ \mu\text{eV}$, $\hbar\kappa = 50 \ \mu\text{eV}$, giving $\frac{4g^2}{\kappa\gamma} = 32$.

In Figure~\ref{fig1} (a-b), we plot the indistinguishability and emitted photon number for different resonant pulse and cavity parameters.
A thorough analysis of how to simultaneously optimize resonantly pulsed single photon sources is given in Ref.~\cite{gustin_pulsed_2018}; however, the main points are that short pulses should be used to minimize two-photon emission via re-excitation, and a high Purcell factor should be used to maximize collection efficiency through the cavity. The criteria for suppression of two-photon emission is usually $\tau_p \ll \frac{1}{F_P \gamma}$, although this criterion can be relaxed to $\tau_p \ll \frac{1}{\kappa}$ due to the dynamical decoupling between the cavity and exciton that occurs during a short pulse. Furthermore, high Purcell factor, high Q cavities increase photon indisinguishability by filtering the phonon sidebands~\cite{ilessmith17,grange17}. For these parameters, simultaneous indistinguishabilities of $>95 \%$ and emitted cavity photon numbers of $>90  \%$ can be achieved for a short pulse of $\tau_p = 2$ ps. Note that in charge neutral QDs, the biexciton state places a lower limit on the pulse width as mentioned in Section~\ref{modelsA}.
\begin{figure}[b]
\centering
\includegraphics[width=1\linewidth]{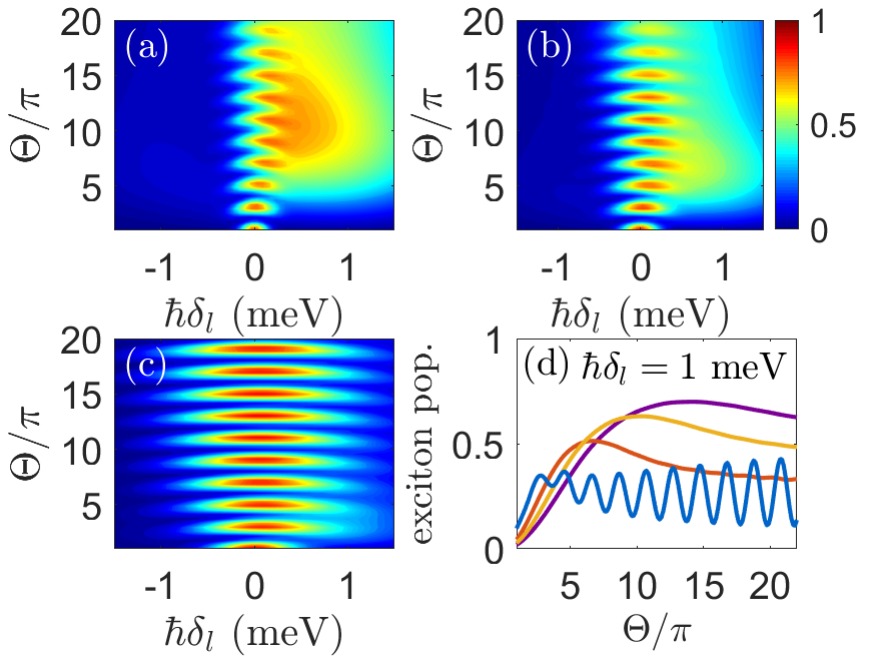}
\caption{\small Exciton population at time $\tau_p$ after the peak of the pulse, for (a) $\tau_p = 6$ ps, (b) $\tau_p = 4$ ps, (c) $\tau_p= 2$ ps, and (d) for pulse widths $\tau_p =2, \, 4, \ 6, \ 8$ ps for blue, red, orange, and purple lines, respectively, at $\hbar\delta_l = 1$ meV.
} 
\label{fig3} 
\end{figure}

\begin{figure}[tbh]
\centering
\includegraphics[width=1\linewidth]{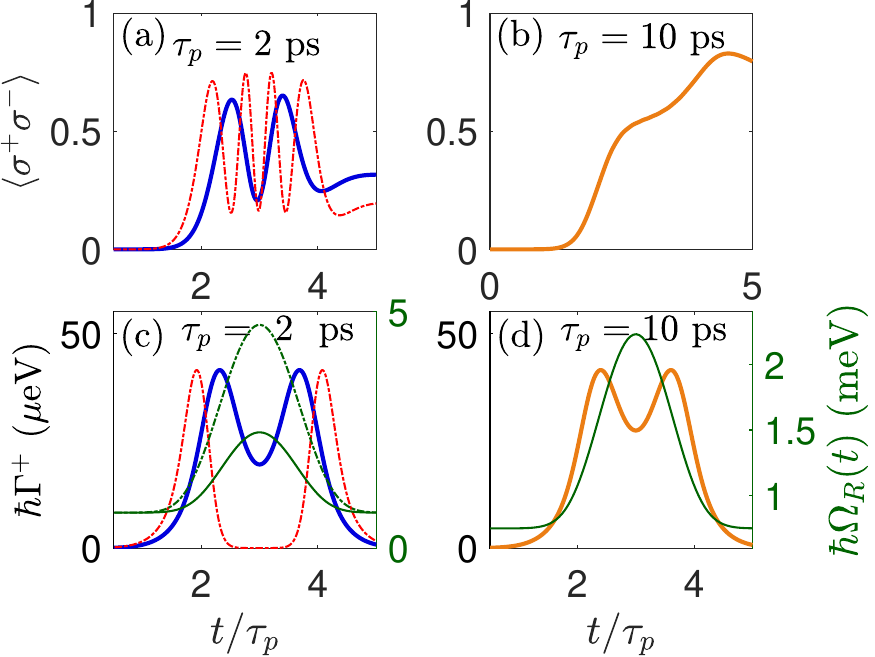}
\caption{\small (a-b) Exciton population and (c-d) phonon scattering rate versus time for different pulse widths. The pulse areas are (d) $18\pi$ (orange solid lines), (c) $4\pi$ (solid lines) and $8\pi$ (dash-dotted lines), and the detuning is $\hbar\delta_l = 750 \ \mu\text{eV}$. For (c-d), the green lines give the value of $\Omega_R(t)$, showing how the $\Gamma^+$ scattering rate dips when the effective drive $\Omega_R$ is greater than a critical value.} 
\label{fig2} 
\end{figure}

Using off-resonant phonon-assisted excitation, one can harness the fact that at low temperatures, phonon emission is much more probable than phonon absorption, and excite with a pulse above the exciton resonance. This leads to an adiabatic preparation of the exciton which is more robust to fluctuations in laser detuning and power. To illustrate this, in Figure~\ref{fig3}, we vary the laser detuning and pulse area and plot the exciton population at a time $2\tau_p$ \emph{after} the peak of the pulse. One can clearly see that as the pulse width is decreased, the efficiency of the inversion process is greatly reduced. Two reasons for this can be seen. In Figure~\ref{fig3}c, it is clear that as the pulse width is reduced, the detuning regime where Rabi oscillations are prominent is increased, as the spectral content of the pulse overlaps with the exciton resonance, interfering with the phonon-assisted inversion process.
Another reason for this can be explained by looking at Figure~\ref{fig2}, in which we plot the exciton populations as a function of time for different pulse widths and pulse areas. We also plot the incoherent excitation  rate
\begin{equation}\label{invrate}
\Gamma^+(t) = \frac{\pi}{4}\frac{\Omega(t)}{\Omega_R(t)}J_p(\Omega_R(t)),
\end{equation}
where $\Omega_R(t) = \sqrt{\Omega^2(t) + \Delta_x^2}$,
which as shown in Appendix~\ref{AppendixB}, is the dominant rate that dictates the phonon-assisted excitation. From Equation~\eqref{invrate}, it is clear that the rate of phonon-assisted excitation will be proportional to the exciton-phonon coupling strength $\alpha$, and will be suppressed for effective drives $\Omega_R \gg \omega_b$. At a pulse width of $\tau_p=10 \ \text{ps}$ (Figure~\ref{fig2}b and \ref{fig2}d), the pulse duration is sufficiently long  such that the system has time to approach the inverted state. Note that the rate $\Gamma^+(t)$ dips in the middle of the pulse, which is due to the effective drive $\Omega_R(t)$ exceeding the spectral response frequency region determined by the phonon spectral function $J_p(\omega)$, and is an indicator of decoupling between the exciton and phonon bath. However, the dip is small enough here to not significantly degrade the inversion process, and in fact partially contributes to the avoidance of  excessive two-photon probabilities by delaying the time-dependent population of the exciton state until the end of the pulse. As we go to lower pulse widths (Figures~\ref{fig2}a and \ref{fig2}c), if we choose a pulse area to \emph{not} allow the phonon bath to strongly decouple during the middle of the pulse, there is not enough time to allow for complete inversion. However, increasing the pulse area to compensate for this will cause strong decoupling of the exciton-phonon system, which also prevents inversion. The relative timescales set by the phonon spectral function parameters thus place a lower bound on the pulse widths that can be used.

On the other hand, the possibility of multi-photon emission places upper limits on the pulse widths that should be used to obtain good single photon figures-of-merit. In Figures~\ref{fig1}c  and \ref{fig1}d, we plot the indistinguishability and emitted photon number against pulse area for two pulse widths, and two different detunings. For a short pulse of $\tau_p = 6 \ \text{ps}$ at $\hbar \delta_l = 0.5 \ \text{meV}$, we have simultaneous $\mathcal{I}\approx 99\%$, and $N_a\approx0.71$, while at a significantly longer pulse of $\tau_p=20 \ \text{ps}$ we have $\mathcal{I}\approx 98\%$, and $N_a \approx 0.9$. This suggests that the off-resonant excitation scheme, in a high-Q cavity system, has the potential to generate single photons of near-unity indistinguishability (very nearly only limited by exciton-cavity dephasing due to intrinsic phonon coupling) and high efficiencies across a broad range of pulse widths, in favourable contrast to the resonantly excited system which is limited, even in a best-case scenario, to $<50\%$ efficiency due to polarization filtering. These findings shed light on a recent work 
which numerically studied the second order coherence of an off-resonantly pumped QD-cavity system~\cite{cosacchi19}. 
We note additionally that this setup can also be effective without using cavity coupling -- for example, with no cavity coupling ($g=0$), $\hbar\delta_l = 0.5 \ \text{meV}$, $\Theta = 20\pi$, and $\tau_p = 15 \ \text{ps}$, we find $\mathcal{I} = 0.95$ and emitted photon number $\gamma\int_0^{\infty}\langle \sigma^+\sigma^-\rangle(t)dt = 0.95$; however, note that this is assuming the phonon sideband is filtered out of the emission spectrum, meaning the actual efficiency would be lower in this no-cavity case (roughly a $\sim 20\%$ cut~\cite{ilessmith17}).


    To help understand the effect of timing jitter on the indistinguishability of the emitted photons from a biexciton-exciton cascade system, as shown in Figure~\ref{fig_s}b, it is useful to consider a simple model of radiative decay from the excited cascade system. Consider the idealization of a pulse which perfectly inverts the system, such that initially the system is in the biexciton state $\rho(t=0) = \ket{u}\bra{u}$. Then, for weakly coupled cavities with  $g/\kappa \ll 1$, the cavity mode can be adiabatically eliminated ($\dot{a}\approx 0$) from the dynamics such that $a \approx -i\frac{2g}{\kappa}\sigma^-_u$ (neglecting trivial input noise terms). This leads to the well-known result that the effective decay from the transition coupled to the cavity mode is enhanced via the Purcell effect $\gamma_u/2 \rightarrow \frac{\gamma_u}{2}(1+F_P)$, with  $F_P = \frac{4g^2}{\kappa\gamma_u/2}$ (note one could alternatively define the Purcell factor to be the enhancement of the total biexciton decay rate). The indistinguishability corresponding to this transition (as well as the exciton-ground transition) is then found to have the simple analytic form:
    \begin{equation}\label{timingj}
    \mathcal{I}= \frac{1}{2}\left[1+\frac{\gamma_u(1+F_P/2)}{\gamma_u(1+F_P/2)+\gamma}\right].
    \end{equation}
    Clearly, then, a high Purcell factor cavity can increase the coherence of the emitted photons for both transitions, by reducing timing jitter.

    Equation~\eqref{timingj} also implies that a cavity with a Purcell enhancement on the exciton to ground transition instead, or a broadband cavity that gives significant Purcell enhancement for both transitions, will lead to poor indistinguishabilities. It is very important to note, however, that this equation was derived with various approximations and is not sufficient for the numerical calculations required for quantitative results. Most importantly, the weak phonon coupling approach misses the broad phonon sidebands that affect the indistinguishability calculation via the first-order coherence function of the exciton-emitted fields. Heuristically, we could expect Equation~\eqref{timingj} to qualitatively describe timing-jitter losses in the indistinguishability of emitted photons if we assume that this phonon sideband is (e.g.) filtered out of emitted photons via post-selection, or through coupling to a cavity with $\kappa \ll \omega_b$; for this work, all numerical results are calculated using the full model described by Equation~\eqref{eq:wpme}, and we present Equation~\eqref{timingj} only for insight into the timing jitter mechanism.

    \begin{figure}[th]
\centering
\includegraphics[width=1\linewidth]{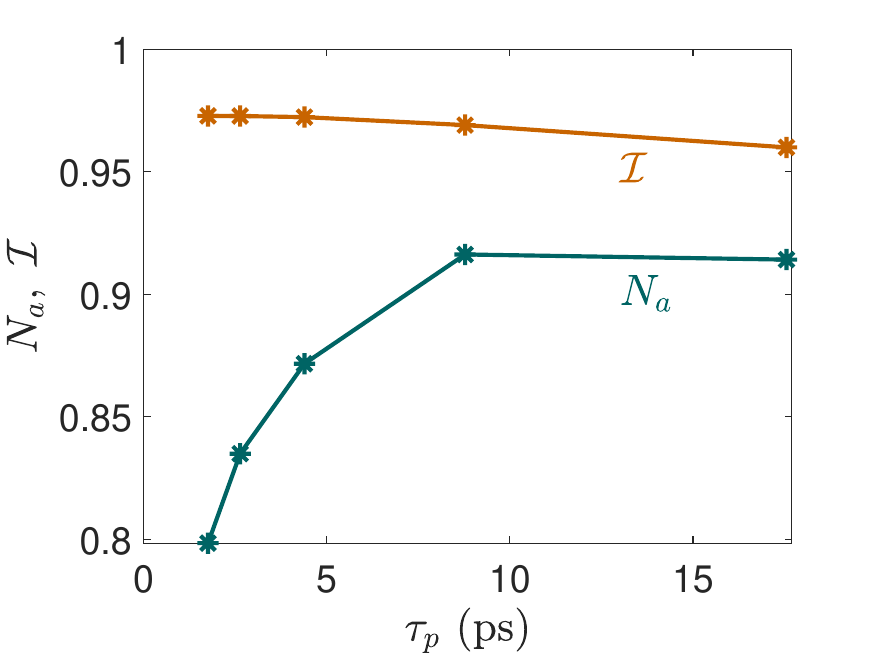}
\caption{\small Biexciton-cavity single photon figures-of-merit for a pulse-triggered SPS via TPE, with a biexciton binding energy of $E_B=3 \ \text{meV}$, and $\gamma_u = 2\gamma$. To achieve an ``effective" two-photon pi-pulse, the pulse area is varied numerically for each simulation.} 

\label{fig4} 
\end{figure}
    
    In Figure~\ref{fig4}, we plot the indistinguishability and emitted cavity photon numbers for different pulse widths. Since the criterion for an ``effective'' pi-pulse when using off-resonant inversion is not strictly defined except in the limit where the intermediate (exciton) state can be adiabatically eliminated from the dynamics---$\hbar\Omega(t) \ll E_B$----we instead vary the pulse area for each simulation to obtain (approximately) the largest emitted cavity photon number, as would be the case in an experiment. 
    As seen in Figure~\ref{fig4}, the indistinguishability remains high for a relatively large range of pulse widths, in contrast to the resonantly excited system. This is due to a high suppression of multiphoton emission~\cite{hanschke18}; i.e., for $\tau_{\text{FWHM}} = 7.3 \ \text{ps}$, we have $D_1=0.02$, but only $D_2 = 3\times10^{-4}$. The indistinguishability is only very slightly lowered ($\sim 0.2\%$) by taking $\gamma_x \rightarrow \gamma_x/2$, $\gamma_u \rightarrow \gamma_u/2$, indicating that the degradation of indistinguishability for this setup and cavity/pulse parameters can thus be attributed almost entirely to intrinsic cavity-induced phonon scattering~\cite{ilessmith17}, as well as (to a smaller extent) timing jitter. As long as the pulse width is large enough as to not excite the exciton ($\frac{\tau_p E_B}{4\hbar} \gg 1$), the figures-of-merit are again somewhat robust against changes in the pulse width (as in the off-resonant phonon-assisted case)---a signature of the adiabatic population inversion processes.

\section{Conclusions}\label{conclusions}
In summary, we have derived
several time-convolutionless MEs to study pulse-driven QD-cavity systems which incorporate electron-phonon scattering rigorously, even for pulses with drive strengths greater than the phonon bath correlation time. We have used these MEs to study three methods of QD-cavity SPS excitation---resonant pi-pulse excitation, off-resonant phonon-assisted inversion, and TPE of the biexciton-exciton cascade.
For our QD-cavity parameters, we find  indistinguishabilities and emitted cavity photon numbers of $>\!95\%$ and $>\!0.9$ simultaneously achievable for a resonant pulse of $\tau_{\text{FWHM}} \approx 3.3 \ \text{ps}$, similar to results of our previous work~\cite{gustin_pulsed_2018}. However, this  resonantly-pumped source requires polarization filtering, reducing the efficiency by at least $50\%$. In contrast, we find for off-resonant phonon-assisted inversion near-unity indistinguishabilities (limited primarily by intrinsic phonon coupling) for a broad range of pulse widths, and for (e.g.) $\tau_{\text{FWHM}} = 33.3 \ \text{ps}$, an emitted photon number of $\sim\!0.9$. Furthermore, this system is more robust to small fluctuations in laser detuning and power. 

We have also investigated via analytical simplification of our ME the role of exciton-phonon decoupling due to strong pulse interactions, which we found to be highly relevant to the dynamics of this scheme both in terms of setting fundamental limits on the pulse widths that can be used, as well as its role in reducing two-photon emission events. Finally, for the SPS based on TPE of the biexciton, we find again near-unity indistinguishabilities, limited primarily by timing jitter and electron-phonon coupling. We have also shown how a high-Purcell factor cavity coupled to the biexciton-exciton transition can reduce dramatically the negative effect of timing jitter on the SPS indistinguishability, and that this system can produce single photons with an ultra-low two-photon probability, as previously shown by Hanschke \emph{et al.}~\cite{hanschke18}. These excitation schemes provide a pathway towards experimental SPSs of even higher efficiency, indistinguishability, and purity, and our intuitive ME approaches provide a simple and insightful framework for modelling such systems.

\section*{Aknowledgements}
This work was supported by the Natural Sciences and Engineering Research Council of Canada (NSERC) and Queen's University. We thank Zheng-Yang Zhou for useful discussions.

\appendix
\section{Comparison of Weak Phonon Coupling with Polaron MEs} \label{AppendixA}    
In the main paper, we use a weak phonon coupling ME to calculate the exciton-phonon coupling via a Born-Markov approximation. However, the validity of the perturbative Born-Markov approximation for describing the interaction can break down at elevated temperatures or for strong phonon coupling parameters~\cite{nazir16}. A polaron transform ME, where one unitarily transforms into a ``polaron'' frame in which the zero-field exciton-phonon coupling is exactly diagonalized before performing the Born-Markov approximation with respect to the field-induced polaron-phonon couplings, has been show to be valid across a much broader range of temperatures and phonon coupling strengths~\cite{mccutcheon11}. However, the polaron transform ME can break down for pulse amplitudes which approach the phonon cutoff frequency $\omega_b$---which is the case for some of the results in Section~\ref{results}~\cite{mccutcheon10}. Thus, in this Appendix we compare the dynamics of a driven QD-cavity system with both MEs in a regime where the pulse amplitude is weak. In this regime, we can expect the polaron transform ME to be rigorously accurate, and as such the regime in which we expect the weak phonon coupling ME to also be valid can be determined by comparison.

    \begin{figure}[b]
\centering
\includegraphics[width=1\linewidth]{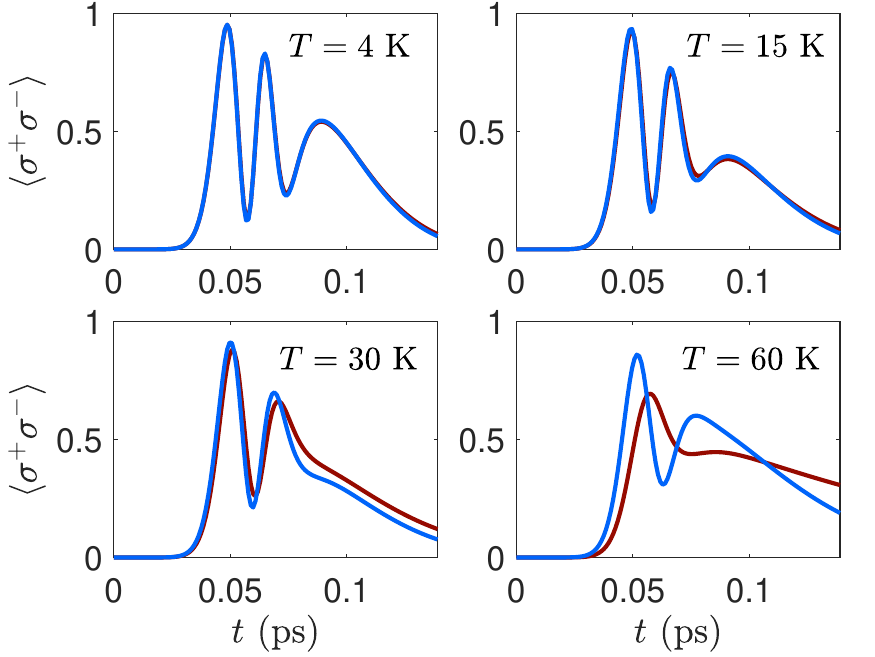}
\caption{\small Exciton populations calculated with the weak-coupling (red) and polaron (blue) MEs for different temperatures. We use $\tau_p = 20 \ \rm{ps}$ and $\Theta = 5\pi$ to ensure that the max pulse amplitude is weak enough such that the polaron approach remains valid. The laser is resonant ($\delta_l =0$).} 
\label{fig1_S} 
\end{figure}

The model used for the weak phonon coupling ME is described by Equation~\eqref{eq:wpme}. The equivalent description of the dynamics via a polaron transform ME is given by (e.g., see Ref.~\cite{gustin_pulsed_2018}):
\begin{equation}\label{mep}
    \frac{\rm{d}\rho}{\rm{dt}} = -\frac{i}{\hbar}[H'_S,\rho] + \frac{1}{2}\sum_{\mu}\mathcal{L}[A_\mu]\rho + \mathbb{L_{\rm{p}}}\rho,
    \end{equation}
    where 
    \begin{align}
      \mathbb{L_{\rm{p}}}\rho= &- \frac{1}{\hbar^2}\int_0^{\infty}d\tau \sum\limits_{\mathclap{m = \{g, u\}}}  \big(G_m(\tau) \times \nonumber \\ &
 [X_m(t), \widetilde{X}_m(t-\tau,t) \rho(t)] + {\rm H.c.} \big),
    \end{align}
    where $X_g = \frac{\hbar\Omega(t)}{2}\sigma_x + \hbar g (\sigma^{+}a + \sigma^{-}a^{\dagger})$, $X_u = -\frac{\hbar\Omega(t)}{2}\sigma_y + i\hbar g(\sigma^+ a - \sigma^- a^\dagger)$.
    Here, $G_g(\tau) = \B^2 (\cosh{(\phi(\tau))}-1)$ and $G_u(\tau) = \B^2\sinh{(\phi(\tau))}$ are the polaron Green functions, with
    \begin{equation}
    \phi(\tau) \!=\! \int\limits_{0}^{\infty}\!d\omega \frac{J_p(\omega)}{\omega^2}\!\left(\coth{\left(\frac{ \hbar \omega}{2 k_B T}\right)}\cos{(\omega\tau)}\! -\! i\sin{(\omega\tau)}\right),
    \end{equation}
    and $\B = e^{-\phi(0)/2}$.
    Within the additional Markov approximation used in Sec~\ref{modelsB}, we have
    \begin{equation}
        \tilde{X}_m(t-\tau,t) \approx e^{-iH'_S(t)\tau/\hbar}X_m(t)e^{iH'_S(t)\tau/\hbar}.
    \end{equation}
    The Lindblad terms are the same as those in the weak phonon coupling ME, and the polaron-renormalized system Hamiltonian $H'_S$ is
    \begin{align}
        H'_S(t) = & -\hbar\delta_l N - \hbar\delta_l a^{\dagger}a \nonumber \\ &+ \frac{\B\hbar\Omega(t)}{2}\sigma_x + \B \hbar g(a\sigma^+ + a^{\dagger}\sigma^-),
    \end{align}
    which is similar to the weak phonon coupling $H_S(t)$, but with the polaron shift included in the system Hamiltonian, and the cavity and pulse strengths being attenuated by a factor $\B$.

    In Figure~\ref{fig1_S}, we compare the dynamics of a driven QD-cavity system using both MEs (Equations~\eqref{eq:wpme} and~\eqref{mep}). While the weak-coupling approach breaks down at high temperatures, for $T=4$ K (and even $T=15$ K)  it remains highly accurate. We note that as we calculate two-time correlation functions using the ME solution via the quantum regression theorem, they can also be expected to agree at low temperatures for both weak phonon coupling and polaron approaches.

\section{Analytical Simplification of the Master Equation}\label{AppendixB}

\begin{figure}[t]
\centering
\includegraphics[width=1\linewidth]{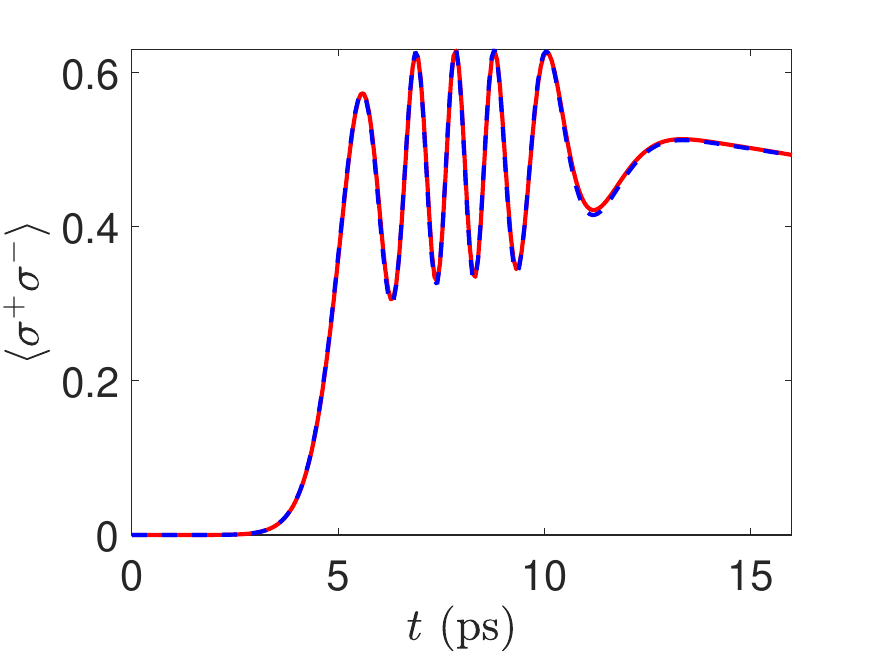}
\caption{\small Comparison of populations calculated with analytically simplified ME of Equation~\eqref{eq:ame} (solid red) versus the ME with the phonon dissipator of Equation~\eqref{eq:me} (dashed blue). 
The differences are, as expected, negligible during the pulse, and as such populations immediately following the pulse can be calculated using the approximate ME for short pulses. Here, $\tau_p = 4 \ \text{ps}$, $\Theta = 10\pi$, and $\hbar\delta_l= 500 \ \mu \text{eV}$. 
}
\label{fig:sim}
\end{figure}

In this Appendix, we give an analytical simplification of Equation~\eqref{eq:me} by
neglecting the influence of the cavity-exciton coupling term of the system 
Hamiltonian in the transformation $\tilde{N}(-\tau) = 
e^{-iH_S(t)\tau/\hbar}Ne^{iH_S(t)\tau/\hbar}$. This approximation is appropriate for 
short pulses, as in the regimes studied here the maximum pulse amplitudes (and detunings in the off-resonant case) are much larger than $g$---see 
Figure~\ref{fig:sim}. We can then commute the remaining cavity term, and the 
remaining transformation acts only on the ground-exciton subspace, through:
\begin{equation}
    \tilde{N}(-\tau) = e^{-A}Ne^{A},
\end{equation} where
\begin{equation}
    A = i\tau\left(\Delta_x N + \frac{\Omega(t)}{2}\sigma_x\right).
\end{equation}
These matrix exponentials are straightforwardly calculated by diagonalizing $A$, and we find
\begin{align}
    \tilde{N}(-\tau) =& N - \frac{\Omega^2(t)}{\Omega_R^2(t)}\sin^2{\left(\frac{\Omega_R(t)\tau}{2}\right)}\sigma_z \nonumber \\ &+ \frac{\Omega(t)\Delta_x}{\Omega_R^2(t)}\sin^2{\left(\frac{\Omega_R(t)\tau}{2}\right)}\sigma_x \nonumber \\ &- \frac{\Omega(t)}{2\Omega_R(t)}\sin{\left(\Omega_R(t)\tau\right)}\sigma_y,
\end{align}
where $\sigma_z = \sigma^+\sigma^- - \sigma^-\sigma^+$. Expanding the ME and using the fact that $\int_0^{\infty} d\tau \Gamma_{\text{w}}(\tau) = -i\Delta_P$, we arrive at the explicit analytical result:
\begin{align}\label{eq:ame}
\mathbb{L}_{\text{w}}\rho &= -i[-\Delta_P N,\rho] + \frac{\gamma'_{\text{eff}}(t)}{2}\mathcal{L}[N]\rho \nonumber \\ & -i\frac{\Omega\Delta_x}{\Omega_R^2}\left[\frac{\Delta_P}{2}\!+\! \text{Im}\{R_c\}\right]\left(\sigma_x\rho N\! -\! N\sigma_x \rho\! -\! \text{H.c.}\right) \nonumber \\ & - \text{Re}\{R_c\}\frac{\Omega\Delta_x}{\Omega_R^2}\left(\sigma_x \rho N - N\sigma_x \rho + \text{H.c.}\right) \nonumber \\ & - \text{Re}\{R_s\}\frac{\Omega}{\Omega_R}\left(\sigma_y \rho N - N\sigma_y \rho + \text{H.c.}\right) \nonumber \\ &-i\text{Im}\{R_s\}\frac{\Omega}{\Omega_R}\left(\sigma_y\rho N-N\sigma_y \rho - \text{H.c.}\right),
\end{align}
where we have suppressed the explicit functional dependence on time of $\Omega(t)$, $\Omega_R(t) = \sqrt{\Omega^2(t) + \Delta_x^2}$, $R_c(t)$, and $R_s(t)$ for notational brevity, and 
\begin{equation}
    R_c(t) = \frac{1}{2}\int_{0}^{\infty} d\tau \Gamma_{\text{w}}(\tau) \cos{\left(\Omega_R(t)\tau\right)},
\end{equation}
\begin{equation}
    R_s(t) = \frac{1}{2}\int_{0}^{\infty} d\tau \Gamma_{\text{w}}(\tau) \sin{\left(\Omega_R(t)\tau\right)}.
\end{equation}
The effective pulse-driven pure dephasing rate is
\begin{align}
    \gamma'_{\text{eff}}(t) &= 4\left(\frac{\Omega(t)}{\Omega_R(t)}\right)^2\text{Re}\{R_c(t)\} \nonumber \\ & = \pi\left(\frac{\Omega(t)}{\Omega_R(t)}\right)^2J_p(\Omega_R(t))\coth{\left(\frac{\Omega_R(t)}{2k_B T}\right)}.
\end{align}

Through careful analysis of, e.g., the Bloch equations formed by considering the matrix elements of Equation~\eqref{eq:ame}, as well as with numerical calculations, 
we can see that for negative detuning $\Delta_x$ (above resonance excitation), the rate corresponding most closely to the phonon-assisted inversion of the exciton is
\begin{equation}
    \Gamma^+(t) = -\text{Im}\{R_s(t)\}\frac{\Omega(t)}{\Omega_R(t)} - \text{Re}\{R_c(t)\}\frac{\Omega(t)\Delta_x}{\Omega_R(t)^2}.
\end{equation}
The second term is, in the regimes studied here, smaller than the first 
and so for the sake of qualitatively gaining better insight we can approximate
\begin{align}
    \Gamma^+(t) &\approx -\text{Im}\{R_s(t)\}\frac{\Omega(t)}{\Omega_R(t)} \nonumber \\ & = \frac{\pi}{4}\frac{\Omega(t)}{\Omega_R(t)}J_p(\Omega_R(t)),
\end{align}
as used in Section~\ref{results}. 

In Figure~\ref{fig:sim}, we compare the population dynamics with this approximate simplified ME versus the full result (see figure caption for precise equations used), which includes the effect of the cavity terms in the $\tilde{N}(-\tau)$ transformation. As expected, we see excellent agreement for a short pulse; note that the cavity terms in the transformation are responsible for cavity-exciton pure dephasing which causes a degradation of the single photon indistinguishability~\cite{ilessmith17}, and as such they should be considered for full simulations of the radiative dynamics. As such, we only use the simplified ME for the results of Figures~\ref{fig3} and~\ref{fig2}.

\bibliography{SPS_Comp_Revise3}

\begin{thebibliography}{41}%
\makeatletter
\providecommand \@ifxundefined [1]{%
 \@ifx{#1\undefined}
}%
\providecommand \@ifnum [1]{%
 \ifnum #1\expandafter \@firstoftwo
 \else \expandafter \@secondoftwo
 \fi
}%
\providecommand \@ifx [1]{%
 \ifx #1\expandafter \@firstoftwo
 \else \expandafter \@secondoftwo
 \fi
}%
\providecommand \natexlab [1]{#1}%
\providecommand \enquote  [1]{``#1''}%
\providecommand \bibnamefont  [1]{#1}%
\providecommand \bibfnamefont [1]{#1}%
\providecommand \citenamefont [1]{#1}%
\providecommand \href@noop [0]{\@secondoftwo}%
\providecommand \href [0]{\begingroup \@sanitize@url \@href}%
\providecommand \@href[1]{\@@startlink{#1}\@@href}%
\providecommand \@@href[1]{\endgroup#1\@@endlink}%
\providecommand \@sanitize@url [0]{\catcode `\\12\catcode `\$12\catcode
  `\&12\catcode `\#12\catcode `\^12\catcode `\_12\catcode `\%12\relax}%
\providecommand \@@startlink[1]{}%
\providecommand \@@endlink[0]{}%
\providecommand \url  [0]{\begingroup\@sanitize@url \@url }%
\providecommand \@url [1]{\endgroup\@href {#1}{\urlprefix }}%
\providecommand \urlprefix  [0]{URL }%
\providecommand \Eprint [0]{\href }%
\providecommand \doibase [0]{http://dx.doi.org/}%
\providecommand \selectlanguage [0]{\@gobble}%
\providecommand \bibinfo  [0]{\@secondoftwo}%
\providecommand \bibfield  [0]{\@secondoftwo}%
\providecommand \translation [1]{[#1]}%
\providecommand \BibitemOpen [0]{}%
\providecommand \bibitemStop [0]{}%
\providecommand \bibitemNoStop [0]{.\EOS\space}%
\providecommand \EOS [0]{\spacefactor3000\relax}%
\providecommand \BibitemShut  [1]{\csname bibitem#1\endcsname}%
\let\auto@bib@innerbib\@empty
\bibitem [{\citenamefont {Kiraz}\ \emph {et~al.}(2004)\citenamefont {Kiraz},
  \citenamefont {Atat{\"u}re},\ and\ \citenamefont {Imamo\v{g}lu}}]{kiraz04}%
  \BibitemOpen
  \bibfield  {author} {\bibinfo {author} {\bibfnamefont {A.}~\bibnamefont
  {Kiraz}}, \bibinfo {author} {\bibfnamefont {M.}~\bibnamefont {Atat{\"u}re}},
  \ and\ \bibinfo {author} {\bibfnamefont {A.}~\bibnamefont {Imamo\v{g}lu}},\
  }\bibfield  {title} {\enquote {\bibinfo {title} {Quantum-dot single-photon
  sources: Prospects for applications in linear optics quantum-information
  processing},}\ }\href@noop {} {\bibfield  {journal} {\bibinfo  {journal}
  {Phys. Rev. A}\ }\textbf {\bibinfo {volume} {69}},\ \bibinfo {pages} {032305}
  (\bibinfo {year} {2004})}\BibitemShut {NoStop}%
\bibitem [{\citenamefont {Michler}\ \emph {et~al.}(2000)\citenamefont
  {Michler}, \citenamefont {Kiraz}, \citenamefont {Schoenfeld}, \citenamefont
  {Petroff}, \citenamefont {Zhang}, \citenamefont {Hu},\ and\ \citenamefont
  {Imamoglu}}]{michler00}%
  \BibitemOpen
  \bibfield  {author} {\bibinfo {author} {\bibfnamefont {P.}~\bibnamefont
  {Michler}}, \bibinfo {author} {\bibfnamefont {A.}~\bibnamefont {Kiraz}},
  \bibinfo {author} {\bibfnamefont {W.~V.}\ \bibnamefont {Schoenfeld}},
  \bibinfo {author} {\bibfnamefont {P.~M.}\ \bibnamefont {Petroff}}, \bibinfo
  {author} {\bibfnamefont {Lidong}\ \bibnamefont {Zhang}}, \bibinfo {author}
  {\bibfnamefont {E.}~\bibnamefont {Hu}}, \ and\ \bibinfo {author}
  {\bibfnamefont {A.}~\bibnamefont {Imamoglu}},\ }\bibfield  {title} {\enquote
  {\bibinfo {title} {A quantum dot single-photon turnstile device},}\
  }\href@noop {} {\bibfield  {journal} {\bibinfo  {journal} {Science}\ }\textbf
  {\bibinfo {volume} {290}},\ \bibinfo {pages} {2282--2285} (\bibinfo {year}
  {2000})}\BibitemShut {NoStop}%
\bibitem [{\citenamefont {Iles-Smith}\ \emph {et~al.}(2017)\citenamefont
  {Iles-Smith}, \citenamefont {McCutcheon}, \citenamefont {Nazir},\ and\
  \citenamefont {M\o{}rk}}]{ilessmith17}%
  \BibitemOpen
  \bibfield  {author} {\bibinfo {author} {\bibfnamefont {J.}~\bibnamefont
  {Iles-Smith}}, \bibinfo {author} {\bibfnamefont {D.~P.~S.}\ \bibnamefont
  {McCutcheon}}, \bibinfo {author} {\bibfnamefont {A.}~\bibnamefont {Nazir}}, \
  and\ \bibinfo {author} {\bibfnamefont {J.}~\bibnamefont {M\o{}rk}},\
  }\bibfield  {title} {\enquote {\bibinfo {title} {Phonon scattering inhibits
  simultaneous near-unity efficiency and indistinguishability in semiconductor
  single-photon sources},}\ }\href@noop {} {\bibfield  {journal} {\bibinfo
  {journal} {Nat. Phot.}\ }\textbf {\bibinfo {volume} {11}},\ \bibinfo {pages}
  {521--526} (\bibinfo {year} {2017})}\BibitemShut {NoStop}%
\bibitem [{\citenamefont {Gustin}\ and\ \citenamefont
  {Hughes}(2017)}]{gustin17}%
  \BibitemOpen
  \bibfield  {author} {\bibinfo {author} {\bibfnamefont {C.}~\bibnamefont
  {Gustin}}\ and\ \bibinfo {author} {\bibfnamefont {S.}~\bibnamefont
  {Hughes}},\ }\bibfield  {title} {\enquote {\bibinfo {title} {Influence of
  electron-phonon scattering for an on-demand quantum dot single-photon source
  using cavity-assisted adiabatic passage},}\ }\href@noop {} {\bibfield
  {journal} {\bibinfo  {journal} {Phys. Rev. B}\ }\textbf {\bibinfo {volume}
  {96}},\ \bibinfo {pages} {085305} (\bibinfo {year} {2017})}\BibitemShut
  {NoStop}%
\bibitem [{\citenamefont {Grange}\ \emph {et~al.}(2017)\citenamefont {Grange},
  \citenamefont {Somaschi}, \citenamefont {Ant\'{o}n}, \citenamefont
  {De~Santis}, \citenamefont {Coppola}, \citenamefont {Giesz}, \citenamefont
  {Lema\^{i}tre}, \citenamefont {Sagnes}, \citenamefont {Auff\'{e}ves},\ and\
  \citenamefont {Senellart}}]{grange17}%
  \BibitemOpen
  \bibfield  {author} {\bibinfo {author} {\bibfnamefont {T.}~\bibnamefont
  {Grange}}, \bibinfo {author} {\bibfnamefont {N.}~\bibnamefont {Somaschi}},
  \bibinfo {author} {\bibfnamefont {C.}~\bibnamefont {Ant\'{o}n}}, \bibinfo
  {author} {\bibfnamefont {L.}~\bibnamefont {De~Santis}}, \bibinfo {author}
  {\bibfnamefont {G.}~\bibnamefont {Coppola}}, \bibinfo {author} {\bibfnamefont
  {V.}~\bibnamefont {Giesz}}, \bibinfo {author} {\bibfnamefont
  {A.}~\bibnamefont {Lema\^{i}tre}}, \bibinfo {author} {\bibfnamefont
  {I.}~\bibnamefont {Sagnes}}, \bibinfo {author} {\bibfnamefont
  {A.}~\bibnamefont {Auff\'{e}ves}}, \ and\ \bibinfo {author} {\bibfnamefont
  {P.}~\bibnamefont {Senellart}},\ }\bibfield  {title} {\enquote {\bibinfo
  {title} {{Reducing Phonon-Induced Decoherence in Solid-State Single-Photon
  Sources with Cavity Quantum Electrodynamics}},}\ }\href@noop {} {\bibfield
  {journal} {\bibinfo  {journal} {Phys. Rev. Lett.}\ }\textbf {\bibinfo
  {volume} {118}},\ \bibinfo {pages} {253602} (\bibinfo {year}
  {2017})}\BibitemShut {NoStop}%
\bibitem [{\citenamefont {Manson}\ \emph {et~al.}(2016)\citenamefont {Manson},
  \citenamefont {Roy-Choudhury},\ and\ \citenamefont {Hughes}}]{ross16}%
  \BibitemOpen
  \bibfield  {author} {\bibinfo {author} {\bibfnamefont {R.}~\bibnamefont
  {Manson}}, \bibinfo {author} {\bibfnamefont {K.}~\bibnamefont
  {Roy-Choudhury}}, \ and\ \bibinfo {author} {\bibfnamefont {S.}~\bibnamefont
  {Hughes}},\ }\bibfield  {title} {\enquote {\bibinfo {title} {Polaron master
  equation theory of pulse-driven phonon phonon-assisted population inversion
  and single-photon emission from quantum-dot excitons},}\ }\href@noop {}
  {\bibfield  {journal} {\bibinfo  {journal} {Phys. Rev. B}\ }\textbf {\bibinfo
  {volume} {93}},\ \bibinfo {pages} {155423} (\bibinfo {year}
  {2016})}\BibitemShut {NoStop}%
\bibitem [{\citenamefont {Tighineanu}\ \emph {et~al.}(2018)\citenamefont
  {Tighineanu}, \citenamefont {Dree{\ss}en}, \citenamefont {Flindt},
  \citenamefont {Lodahl},\ and\ \citenamefont {S{\o}rensen}}]{tighineanu18}%
  \BibitemOpen
  \bibfield  {author} {\bibinfo {author} {\bibfnamefont {P.}~\bibnamefont
  {Tighineanu}}, \bibinfo {author} {\bibfnamefont {C.~L.}\ \bibnamefont
  {Dree{\ss}en}}, \bibinfo {author} {\bibfnamefont {C.}~\bibnamefont {Flindt}},
  \bibinfo {author} {\bibfnamefont {P.}~\bibnamefont {Lodahl}}, \ and\ \bibinfo
  {author} {\bibfnamefont {A.~S.}\ \bibnamefont {S{\o}rensen}},\ }\bibfield
  {title} {\enquote {\bibinfo {title} {{Phonon Decoherence of Quantum Dots in
  Photonic Structures: Broadening of the Zero-Phonon Line and the Role of
  Dimensionality}},}\ }\href@noop {} {\bibfield  {journal} {\bibinfo  {journal}
  {Phys. Rev. Lett.}\ }\textbf {\bibinfo {volume} {120}},\ \bibinfo {pages}
  {257401} (\bibinfo {year} {2018})}\BibitemShut {NoStop}%
\bibitem [{\citenamefont {Denning}\ \emph {et~al.}(2018)\citenamefont
  {Denning}, \citenamefont {Iles-Smith}, \citenamefont {Osterkryger},
  \citenamefont {Gregersen},\ and\ \citenamefont {Mork}}]{denning18}%
  \BibitemOpen
  \bibfield  {author} {\bibinfo {author} {\bibfnamefont {Emil~V.}\ \bibnamefont
  {Denning}}, \bibinfo {author} {\bibfnamefont {Jake}\ \bibnamefont
  {Iles-Smith}}, \bibinfo {author} {\bibfnamefont {Andreas~Dyhl}\ \bibnamefont
  {Osterkryger}}, \bibinfo {author} {\bibfnamefont {Niels}\ \bibnamefont
  {Gregersen}}, \ and\ \bibinfo {author} {\bibfnamefont {Jesper}\ \bibnamefont
  {Mork}},\ }\bibfield  {title} {\enquote {\bibinfo {title} {Cavity-waveguide
  interplay in optical resonators and its role in optimal single-photon
  sources},}\ }\href@noop {} {\bibfield  {journal} {\bibinfo  {journal} {Phys.
  Rev. B}\ }\textbf {\bibinfo {volume} {98}},\ \bibinfo {pages} {121306(R)}
  (\bibinfo {year} {2018})}\BibitemShut {NoStop}%
\bibitem [{\citenamefont {Gerhardt}\ \emph {et~al.}(2018)\citenamefont
  {Gerhardt}, \citenamefont {Iles-Smith}, \citenamefont {McCutcheon},
  \citenamefont {He}, \citenamefont {Unsleber}, \citenamefont {Betzold},
  \citenamefont {Gregersen}, \citenamefont {M{\o}rk}, \citenamefont
  {H\"{o}fling},\ and\ \citenamefont {Schneider}}]{gerhardt18}%
  \BibitemOpen
  \bibfield  {author} {\bibinfo {author} {\bibfnamefont {Stefan}\ \bibnamefont
  {Gerhardt}}, \bibinfo {author} {\bibfnamefont {Jake}\ \bibnamefont
  {Iles-Smith}}, \bibinfo {author} {\bibfnamefont {Dara P.~S.}\ \bibnamefont
  {McCutcheon}}, \bibinfo {author} {\bibfnamefont {Yu-Ming}\ \bibnamefont
  {He}}, \bibinfo {author} {\bibfnamefont {Sebastian}\ \bibnamefont
  {Unsleber}}, \bibinfo {author} {\bibfnamefont {Simon}\ \bibnamefont
  {Betzold}}, \bibinfo {author} {\bibfnamefont {Niels}\ \bibnamefont
  {Gregersen}}, \bibinfo {author} {\bibfnamefont {Jesper}\ \bibnamefont
  {M{\o}rk}}, \bibinfo {author} {\bibfnamefont {Sven}\ \bibnamefont
  {H\"{o}fling}}, \ and\ \bibinfo {author} {\bibfnamefont {Christian}\
  \bibnamefont {Schneider}},\ }\bibfield  {title} {\enquote {\bibinfo {title}
  {Intrinsic and environmental effects on the interference properties of a
  high-performance quantum dot single-photon source},}\ }\href@noop {}
  {\bibfield  {journal} {\bibinfo  {journal} {Phys. Rev. B}\ }\textbf {\bibinfo
  {volume} {97}},\ \bibinfo {pages} {195432} (\bibinfo {year}
  {2018})}\BibitemShut {NoStop}%
\bibitem [{\citenamefont {Gustin}\ and\ \citenamefont
  {Hughes}(2018)}]{gustin_pulsed_2018}%
  \BibitemOpen
  \bibfield  {author} {\bibinfo {author} {\bibfnamefont {Chris}\ \bibnamefont
  {Gustin}}\ and\ \bibinfo {author} {\bibfnamefont {Stephen}\ \bibnamefont
  {Hughes}},\ }\bibfield  {title} {\enquote {\bibinfo {title} {Pulsed
  excitation dynamics in quantum-dot{\textendash}cavity systems: Limits to
  optimizing the fidelity of on-demand single-photon sources},}\ }\href@noop {}
  {\bibfield  {journal} {\bibinfo  {journal} {Phys. Rev. B}\ }\textbf {\bibinfo
  {volume} {98}} (\bibinfo {year} {2018})}\BibitemShut {NoStop}%
\bibitem [{\citenamefont {Fischer}\ \emph
  {et~al.}(2017{\natexlab{a}})\citenamefont {Fischer}, \citenamefont
  {Hanschke}, \citenamefont {Kremser}, \citenamefont {Finley}, \citenamefont
  {M{\"u}ller},\ and\ \citenamefont {Vu\v{c}kovi\'{c}}}]{fischer17}%
  \BibitemOpen
  \bibfield  {author} {\bibinfo {author} {\bibfnamefont {K.~A.}\ \bibnamefont
  {Fischer}}, \bibinfo {author} {\bibfnamefont {L.}~\bibnamefont {Hanschke}},
  \bibinfo {author} {\bibfnamefont {M.}~\bibnamefont {Kremser}}, \bibinfo
  {author} {\bibfnamefont {J.~J.}\ \bibnamefont {Finley}}, \bibinfo {author}
  {\bibfnamefont {K.}~\bibnamefont {M{\"u}ller}}, \ and\ \bibinfo {author}
  {\bibfnamefont {J.}~\bibnamefont {Vu\v{c}kovi\'{c}}},\ }\bibfield  {title}
  {\enquote {\bibinfo {title} {{Pulsed Rabi oscillations in quantum two-level
  systems: beyond the Area Theorem}},}\ }\href@noop {} {\bibfield  {journal}
  {\bibinfo  {journal} {Quantum Sci. Technol.}\ }\textbf {\bibinfo {volume}
  {3}},\ \bibinfo {pages} {014006} (\bibinfo {year}
  {2017}{\natexlab{a}})}\BibitemShut {NoStop}%
\bibitem [{\citenamefont {Fischer}\ \emph
  {et~al.}(2017{\natexlab{b}})\citenamefont {Fischer}, \citenamefont
  {Hanschke}, \citenamefont {Wierzbowski}, \citenamefont {Simmet},
  \citenamefont {Dory}, \citenamefont {Finley}, \citenamefont
  {Vu\v{c}kovi\'{c}},\ and\ \citenamefont {M{\"u}ller}}]{2fischer17}%
  \BibitemOpen
  \bibfield  {author} {\bibinfo {author} {\bibfnamefont {K.~A.}\ \bibnamefont
  {Fischer}}, \bibinfo {author} {\bibfnamefont {L.}~\bibnamefont {Hanschke}},
  \bibinfo {author} {\bibfnamefont {J.}~\bibnamefont {Wierzbowski}}, \bibinfo
  {author} {\bibfnamefont {T.}~\bibnamefont {Simmet}}, \bibinfo {author}
  {\bibfnamefont {C.}~\bibnamefont {Dory}}, \bibinfo {author} {\bibfnamefont
  {J.~J.}\ \bibnamefont {Finley}}, \bibinfo {author} {\bibfnamefont
  {J.}~\bibnamefont {Vu\v{c}kovi\'{c}}}, \ and\ \bibinfo {author}
  {\bibfnamefont {K.}~\bibnamefont {M{\"u}ller}},\ }\bibfield  {title}
  {\enquote {\bibinfo {title} {Signatures of two-photon pulses from a quantum
  two-level system},}\ }\href@noop {} {\bibfield  {journal} {\bibinfo
  {journal} {Nat. Phys.}\ }\textbf {\bibinfo {volume} {13}},\ \bibinfo {pages}
  {649--654} (\bibinfo {year} {2017}{\natexlab{b}})}\BibitemShut {NoStop}%
\bibitem [{\citenamefont {Reed}\ \emph {et~al.}(2016)\citenamefont {Reed} \emph
  {et~al.}}]{reed16}%
  \BibitemOpen
  \bibfield  {author} {\bibinfo {author} {\bibfnamefont {M.~D.}\ \bibnamefont
  {Reed}} \emph {et~al.},\ }\bibfield  {title} {\enquote {\bibinfo {title}
  {{Reduced Sensitivity to Charge Noise in Semiconductor Spin Qubits via
  Symmetric Operation}},}\ }\href@noop {} {\bibfield  {journal} {\bibinfo
  {journal} {Phys. Rev. Lett.}\ }\textbf {\bibinfo {volume} {116}},\ \bibinfo
  {pages} {110402} (\bibinfo {year} {2016})}\BibitemShut {NoStop}%
\bibitem [{\citenamefont {Kuhlmann}\ \emph {et~al.}(2013)\citenamefont
  {Kuhlmann}, \citenamefont {Houel}, \citenamefont {Ludwig}, \citenamefont
  {Greuter}, \citenamefont {Reuter}, \citenamefont {Wieck}, \citenamefont
  {Poggio},\ and\ \citenamefont {Warburton}}]{kuhlmann13}%
  \BibitemOpen
  \bibfield  {author} {\bibinfo {author} {\bibfnamefont {A.~V.}\ \bibnamefont
  {Kuhlmann}}, \bibinfo {author} {\bibfnamefont {J.}~\bibnamefont {Houel}},
  \bibinfo {author} {\bibfnamefont {A.}~\bibnamefont {Ludwig}}, \bibinfo
  {author} {\bibfnamefont {L.}~\bibnamefont {Greuter}}, \bibinfo {author}
  {\bibfnamefont {D.}~\bibnamefont {Reuter}}, \bibinfo {author} {\bibfnamefont
  {A.~D.}\ \bibnamefont {Wieck}}, \bibinfo {author} {\bibfnamefont
  {M.}~\bibnamefont {Poggio}}, \ and\ \bibinfo {author} {\bibfnamefont {R.~J.}\
  \bibnamefont {Warburton}},\ }\bibfield  {title} {\enquote {\bibinfo {title}
  {Charge noise and spin noise in a semiconductor quantum device},}\
  }\href@noop {} {\bibfield  {journal} {\bibinfo  {journal} {Nat. Phys.}\
  }\textbf {\bibinfo {volume} {9}},\ \bibinfo {pages} {570} (\bibinfo {year}
  {2013})}\BibitemShut {NoStop}%
\bibitem [{\citenamefont {Somaschi}\ \emph {et~al.}(2016)\citenamefont
  {Somaschi}, \citenamefont {Giesz}, \citenamefont {De~Santis}, \citenamefont
  {Loredo}, \citenamefont {Almeida}, \citenamefont {Hornecker}, \citenamefont
  {Portalupi}, \citenamefont {Grange}, \citenamefont {Ant\'{o}n}, \citenamefont
  {Demory}, \citenamefont {G\'{o}mez}, \citenamefont {Sagnes}, \citenamefont
  {Lanzillotti-Kimura}, \citenamefont {Lema\^{i}tre}, \citenamefont {Auffeves},
  \citenamefont {White}, \citenamefont {Lanco},\ and\ \citenamefont
  {Senellart}}]{somaschi16}%
  \BibitemOpen
  \bibfield  {author} {\bibinfo {author} {\bibfnamefont {N.}~\bibnamefont
  {Somaschi}}, \bibinfo {author} {\bibfnamefont {V.}~\bibnamefont {Giesz}},
  \bibinfo {author} {\bibfnamefont {L.}~\bibnamefont {De~Santis}}, \bibinfo
  {author} {\bibfnamefont {J.~C.}\ \bibnamefont {Loredo}}, \bibinfo {author}
  {\bibfnamefont {M.~P.}\ \bibnamefont {Almeida}}, \bibinfo {author}
  {\bibfnamefont {G.}~\bibnamefont {Hornecker}}, \bibinfo {author}
  {\bibfnamefont {S.~L.}\ \bibnamefont {Portalupi}}, \bibinfo {author}
  {\bibfnamefont {T.}~\bibnamefont {Grange}}, \bibinfo {author} {\bibfnamefont
  {C.}~\bibnamefont {Ant\'{o}n}}, \bibinfo {author} {\bibfnamefont
  {J.}~\bibnamefont {Demory}}, \bibinfo {author} {\bibfnamefont
  {C.}~\bibnamefont {G\'{o}mez}}, \bibinfo {author} {\bibfnamefont
  {I.}~\bibnamefont {Sagnes}}, \bibinfo {author} {\bibfnamefont {N.~D.}\
  \bibnamefont {Lanzillotti-Kimura}}, \bibinfo {author} {\bibfnamefont
  {A.}~\bibnamefont {Lema\^{i}tre}}, \bibinfo {author} {\bibfnamefont
  {A.}~\bibnamefont {Auffeves}}, \bibinfo {author} {\bibfnamefont {A.~G.}\
  \bibnamefont {White}}, \bibinfo {author} {\bibfnamefont {L.}~\bibnamefont
  {Lanco}}, \ and\ \bibinfo {author} {\bibfnamefont {P.}~\bibnamefont
  {Senellart}},\ }\bibfield  {title} {\enquote {\bibinfo {title} {Near-optimal
  single-photon sources in the solid state},}\ }\href@noop {} {\bibfield
  {journal} {\bibinfo  {journal} {Nat. Photon.}\ }\textbf {\bibinfo {volume}
  {10}},\ \bibinfo {pages} {340--345} (\bibinfo {year} {2016})}\BibitemShut
  {NoStop}%
\bibitem [{\citenamefont {Senellart}\ \emph {et~al.}(2017)\citenamefont
  {Senellart}, \citenamefont {Solomon},\ and\ \citenamefont
  {White}}]{senellart17}%
  \BibitemOpen
  \bibfield  {author} {\bibinfo {author} {\bibfnamefont {Pascale}\ \bibnamefont
  {Senellart}}, \bibinfo {author} {\bibfnamefont {Glenn}\ \bibnamefont
  {Solomon}}, \ and\ \bibinfo {author} {\bibfnamefont {Andrew}\ \bibnamefont
  {White}},\ }\bibfield  {title} {\enquote {\bibinfo {title} {High-performance
  semiconductor quantum-dot single-photon sources},}\ }\href@noop {} {\bibfield
   {journal} {\bibinfo  {journal} {Nat. Nanotech.}\ }\textbf {\bibinfo {volume}
  {12}},\ \bibinfo {pages} {1026--1039} (\bibinfo {year} {2017})}\BibitemShut
  {NoStop}%
\bibitem [{\citenamefont {{Ding, X. and others}}(2016)}]{ding16}%
  \BibitemOpen
  \bibfield  {author} {\bibinfo {author} {\bibnamefont {{Ding, X. and
  others}}},\ }\bibfield  {title} {\enquote {\bibinfo {title} {{On-Demand
  Single Photons with High Extraction Efficiency and Near-Unity
  Indistinguishability from a Resonantly Driven Quantum Dot in a
  Micropillar}},}\ }\href@noop {} {\bibfield  {journal} {\bibinfo  {journal}
  {Phys. Rev. Lett.}\ }\textbf {\bibinfo {volume} {116}},\ \bibinfo {pages}
  {020401} (\bibinfo {year} {2016})}\BibitemShut {NoStop}%
\bibitem [{\citenamefont {Hanschke}\ \emph {et~al.}(2018)\citenamefont
  {Hanschke}, \citenamefont {Fischer}, \citenamefont {Appel}, \citenamefont
  {Lukin}, \citenamefont {J.}, \citenamefont {Sun}, \citenamefont {Trivedi},
  \citenamefont {Vu\v{c}kovi\'{c}}, \citenamefont {Finley},\ and\ \citenamefont
  {M\"{u}ller}}]{hanschke18}%
  \BibitemOpen
  \bibfield  {author} {\bibinfo {author} {\bibfnamefont {L.}~\bibnamefont
  {Hanschke}}, \bibinfo {author} {\bibfnamefont {K.~A.}\ \bibnamefont
  {Fischer}}, \bibinfo {author} {\bibfnamefont {S.}~\bibnamefont {Appel}},
  \bibinfo {author} {\bibfnamefont {D.}~\bibnamefont {Lukin}}, \bibinfo
  {author} {\bibfnamefont {Wierzbowski}\ \bibnamefont {J.}}, \bibinfo {author}
  {\bibfnamefont {S.}~\bibnamefont {Sun}}, \bibinfo {author} {\bibfnamefont
  {R.}~\bibnamefont {Trivedi}}, \bibinfo {author} {\bibfnamefont
  {J.}~\bibnamefont {Vu\v{c}kovi\'{c}}}, \bibinfo {author} {\bibfnamefont
  {J.~F.}\ \bibnamefont {Finley}}, \ and\ \bibinfo {author} {\bibfnamefont
  {K.}~\bibnamefont {M\"{u}ller}},\ }\bibfield  {title} {\enquote {\bibinfo
  {title} {Quantum dot single-photon sources with ultra-low multi-photon
  probability},}\ }\href@noop {} {\bibfield  {journal} {\bibinfo  {journal}
  {npj Quantum Inf.}\ }\textbf {\bibinfo {volume} {4}},\ \bibinfo {pages} {43}
  (\bibinfo {year} {2018})}\BibitemShut {NoStop}%
\bibitem [{\citenamefont {Schweickert}\ \emph {et~al.}(2018)\citenamefont
  {Schweickert}, \citenamefont {J\"{o}ns}, \citenamefont {Zeuner},
  \citenamefont {Covre~da Silva}, \citenamefont {Huang}, \citenamefont
  {Lettner}, \citenamefont {Reindl}, \citenamefont {Zichi}, \citenamefont
  {Trotta}, \citenamefont {Rastelli},\ and\ \citenamefont
  {Zwiller}}]{schweickert18}%
  \BibitemOpen
  \bibfield  {author} {\bibinfo {author} {\bibfnamefont {Lucas}\ \bibnamefont
  {Schweickert}}, \bibinfo {author} {\bibfnamefont {Klaus~D.}\ \bibnamefont
  {J\"{o}ns}}, \bibinfo {author} {\bibfnamefont {Katharina~D.}\ \bibnamefont
  {Zeuner}}, \bibinfo {author} {\bibfnamefont {Saimon~Filipe}\ \bibnamefont
  {Covre~da Silva}}, \bibinfo {author} {\bibfnamefont {Huiying}\ \bibnamefont
  {Huang}}, \bibinfo {author} {\bibfnamefont {Thomas}\ \bibnamefont {Lettner}},
  \bibinfo {author} {\bibfnamefont {Marcus}\ \bibnamefont {Reindl}}, \bibinfo
  {author} {\bibfnamefont {Julien}\ \bibnamefont {Zichi}}, \bibinfo {author}
  {\bibfnamefont {Rinaldo}\ \bibnamefont {Trotta}}, \bibinfo {author}
  {\bibfnamefont {Armando}\ \bibnamefont {Rastelli}}, \ and\ \bibinfo {author}
  {\bibfnamefont {Val}\ \bibnamefont {Zwiller}},\ }\bibfield  {title} {\enquote
  {\bibinfo {title} {On-demand generation of background-free single photons
  from a solid-state source},}\ }\href@noop {} {\bibfield  {journal} {\bibinfo
  {journal} {Appl. Phys. Lett.}\ }\textbf {\bibinfo {volume} {112}},\ \bibinfo
  {pages} {093106} (\bibinfo {year} {2018})}\BibitemShut {NoStop}%
\bibitem [{\citenamefont {Dusanowski}\ \emph {et~al.}(2019)\citenamefont
  {Dusanowski}, \citenamefont {Kwon}, \citenamefont {Schneider},\ and\
  \citenamefont {H\"{o}fling}}]{dusanowski19}%
  \BibitemOpen
  \bibfield  {author} {\bibinfo {author} {\bibfnamefont {{\L}ukasz}\
  \bibnamefont {Dusanowski}}, \bibinfo {author} {\bibfnamefont {Soon-Hong}\
  \bibnamefont {Kwon}}, \bibinfo {author} {\bibfnamefont {Christian}\
  \bibnamefont {Schneider}}, \ and\ \bibinfo {author} {\bibfnamefont {Sven}\
  \bibnamefont {H\"{o}fling}},\ }\bibfield  {title} {\enquote {\bibinfo {title}
  {{Near-Unity Indistinguishability Single Photon Source for Large-Scale
  Integrated Quantum Optics}},}\ }\href@noop {} {\bibfield  {journal} {\bibinfo
   {journal} {Phys. Rev. Lett.}\ }\textbf {\bibinfo {volume} {122}},\ \bibinfo
  {pages} {173602} (\bibinfo {year} {2019})}\BibitemShut {NoStop}%
\bibitem [{\citenamefont {Liu}\ \emph {et~al.}(2018)\citenamefont {Liu},
  \citenamefont {Brash}, \citenamefont {O'Hara}, \citenamefont {Martins},
  \citenamefont {Phillips}, \citenamefont {Coles}, \citenamefont {Royall},
  \citenamefont {Clarke}, \citenamefont {Bentham}, \citenamefont {Prtljaga},
  \citenamefont {Itskevich}, \citenamefont {Wilson}, \citenamefont {Skolnick},\
  and\ \citenamefont {Fox}}]{liu18}%
  \BibitemOpen
  \bibfield  {author} {\bibinfo {author} {\bibfnamefont {Feng}\ \bibnamefont
  {Liu}}, \bibinfo {author} {\bibfnamefont {Alistair~J.}\ \bibnamefont
  {Brash}}, \bibinfo {author} {\bibfnamefont {John}\ \bibnamefont {O'Hara}},
  \bibinfo {author} {\bibfnamefont {Luis M. P.~P.}\ \bibnamefont {Martins}},
  \bibinfo {author} {\bibfnamefont {Catherine~L.}\ \bibnamefont {Phillips}},
  \bibinfo {author} {\bibfnamefont {Rikki~J.}\ \bibnamefont {Coles}}, \bibinfo
  {author} {\bibfnamefont {Benjamin}\ \bibnamefont {Royall}}, \bibinfo {author}
  {\bibfnamefont {Edmund}\ \bibnamefont {Clarke}}, \bibinfo {author}
  {\bibfnamefont {Christopher}\ \bibnamefont {Bentham}}, \bibinfo {author}
  {\bibfnamefont {Nikola}\ \bibnamefont {Prtljaga}}, \bibinfo {author}
  {\bibfnamefont {Igor~E.}\ \bibnamefont {Itskevich}}, \bibinfo {author}
  {\bibfnamefont {Luke~R.}\ \bibnamefont {Wilson}}, \bibinfo {author}
  {\bibfnamefont {Maurice~S.}\ \bibnamefont {Skolnick}}, \ and\ \bibinfo
  {author} {\bibfnamefont {A.~Mark}\ \bibnamefont {Fox}},\ }\bibfield  {title}
  {\enquote {\bibinfo {title} {{High Purcell factor generation of
  indistinguishable on-chip single photons}},}\ }\href@noop {} {\bibfield
  {journal} {\bibinfo  {journal} {Nat. Nanotech.}\ }\textbf {\bibinfo {volume}
  {13}},\ \bibinfo {pages} {835--840} (\bibinfo {year} {2018})}\BibitemShut
  {NoStop}%
\bibitem [{\citenamefont {Daveau}\ \emph {et~al.}(2017)\citenamefont {Daveau},
  \citenamefont {Balram}, \citenamefont {Pregnolato}, \citenamefont {Liu},
  \citenamefont {Lee}, \citenamefont {Song}, \citenamefont {Verma},
  \citenamefont {Mirin}, \citenamefont {Nam}, \citenamefont {Midolo},
  \citenamefont {Stobbe}, \citenamefont {Srinivasan},\ and\ \citenamefont
  {Lodahl}}]{daveau17}%
  \BibitemOpen
  \bibfield  {author} {\bibinfo {author} {\bibfnamefont {Rapha\"{e}l~S.}\
  \bibnamefont {Daveau}}, \bibinfo {author} {\bibfnamefont {Krishna~C.}\
  \bibnamefont {Balram}}, \bibinfo {author} {\bibfnamefont {Tommaso}\
  \bibnamefont {Pregnolato}}, \bibinfo {author} {\bibfnamefont {Jin}\
  \bibnamefont {Liu}}, \bibinfo {author} {\bibfnamefont {Eun~H.}\ \bibnamefont
  {Lee}}, \bibinfo {author} {\bibfnamefont {Jin~D.}\ \bibnamefont {Song}},
  \bibinfo {author} {\bibfnamefont {Varun}\ \bibnamefont {Verma}}, \bibinfo
  {author} {\bibfnamefont {Richard}\ \bibnamefont {Mirin}}, \bibinfo {author}
  {\bibfnamefont {Sae~Woo}\ \bibnamefont {Nam}}, \bibinfo {author}
  {\bibfnamefont {Leonardo}\ \bibnamefont {Midolo}}, \bibinfo {author}
  {\bibfnamefont {S{\o}ren}\ \bibnamefont {Stobbe}}, \bibinfo {author}
  {\bibfnamefont {Kartik}\ \bibnamefont {Srinivasan}}, \ and\ \bibinfo {author}
  {\bibfnamefont {Peter}\ \bibnamefont {Lodahl}},\ }\bibfield  {title}
  {\enquote {\bibinfo {title} {{Efficient fiber-coupled single-photon source
  based on quantum dots in a photonic-crystal waveguide}},}\ }\href@noop {}
  {\bibfield  {journal} {\bibinfo  {journal} {Optica}\ }\textbf {\bibinfo
  {volume} {4}},\ \bibinfo {pages} {178--184} (\bibinfo {year}
  {2017})}\BibitemShut {NoStop}%
\bibitem [{\citenamefont {Zadeh}\ \emph {et~al.}(2016)\citenamefont {Zadeh},
  \citenamefont {Elshaari}, \citenamefont {J\"{o}ns}, \citenamefont {Fognini},
  \citenamefont {Dalacu}, \citenamefont {Poole}, \citenamefont {Reimer},\ and\
  \citenamefont {Zwiller}}]{zadeh16}%
  \BibitemOpen
  \bibfield  {author} {\bibinfo {author} {\bibfnamefont {Iman~Esmaeil}\
  \bibnamefont {Zadeh}}, \bibinfo {author} {\bibfnamefont {Ali~W.}\
  \bibnamefont {Elshaari}}, \bibinfo {author} {\bibfnamefont {Klaus~D.}\
  \bibnamefont {J\"{o}ns}}, \bibinfo {author} {\bibfnamefont {Andreas}\
  \bibnamefont {Fognini}}, \bibinfo {author} {\bibfnamefont {Dan}\ \bibnamefont
  {Dalacu}}, \bibinfo {author} {\bibfnamefont {Philip~J.}\ \bibnamefont
  {Poole}}, \bibinfo {author} {\bibfnamefont {Michael~E.}\ \bibnamefont
  {Reimer}}, \ and\ \bibinfo {author} {\bibfnamefont {Val}\ \bibnamefont
  {Zwiller}},\ }\bibfield  {title} {\enquote {\bibinfo {title} {{Deterministic
  Integration of Single Photon Sources in Silicon Based Photonic Circuits}},}\
  }\href@noop {} {\bibfield  {journal} {\bibinfo  {journal} {Nano Lett.}\
  }\textbf {\bibinfo {volume} {16}},\ \bibinfo {pages} {2289--2294} (\bibinfo
  {year} {2016})}\BibitemShut {NoStop}%
\bibitem [{\citenamefont {Laucht}\ \emph {et~al.}(2012)\citenamefont {Laucht},
  \citenamefont {P\"{u}tz}, \citenamefont {G\"{u}nthner}, \citenamefont
  {Hauke}, \citenamefont {Saive}, \citenamefont {Fr\'{e}d\'{e}rick},
  \citenamefont {Bichler}, \citenamefont {Amann}, \citenamefont {Holleitner},
  \citenamefont {Kaniber},\ and\ \citenamefont {Finley}}]{laucht12}%
  \BibitemOpen
  \bibfield  {author} {\bibinfo {author} {\bibfnamefont {A.}~\bibnamefont
  {Laucht}}, \bibinfo {author} {\bibfnamefont {S.}~\bibnamefont {P\"{u}tz}},
  \bibinfo {author} {\bibfnamefont {T.}~\bibnamefont {G\"{u}nthner}}, \bibinfo
  {author} {\bibfnamefont {N.}~\bibnamefont {Hauke}}, \bibinfo {author}
  {\bibfnamefont {R.}~\bibnamefont {Saive}}, \bibinfo {author} {\bibfnamefont
  {S.}~\bibnamefont {Fr\'{e}d\'{e}rick}}, \bibinfo {author} {\bibfnamefont
  {M.}~\bibnamefont {Bichler}}, \bibinfo {author} {\bibfnamefont {M.-C.}\
  \bibnamefont {Amann}}, \bibinfo {author} {\bibfnamefont {A.~W.}\ \bibnamefont
  {Holleitner}}, \bibinfo {author} {\bibfnamefont {M.}~\bibnamefont {Kaniber}},
  \ and\ \bibinfo {author} {\bibfnamefont {J.~J.}\ \bibnamefont {Finley}},\
  }\bibfield  {title} {\enquote {\bibinfo {title} {{A Waveguide-Coupled On-Chip
  Single-Photon Source}},}\ }\href@noop {} {\bibfield  {journal} {\bibinfo
  {journal} {Phys. Rev. X}\ }\textbf {\bibinfo {volume} {2}},\ \bibinfo {pages}
  {011014} (\bibinfo {year} {2012})}\BibitemShut {NoStop}%
\bibitem [{\citenamefont {Yao}\ \emph {et~al.}(2010)\citenamefont {Yao},
  \citenamefont {Rao},\ and\ \citenamefont {Hughes}}]{yao10}%
  \BibitemOpen
  \bibfield  {author} {\bibinfo {author} {\bibfnamefont {P.}~\bibnamefont
  {Yao}}, \bibinfo {author} {\bibfnamefont {V.~S. C.~Manga}\ \bibnamefont
  {Rao}}, \ and\ \bibinfo {author} {\bibfnamefont {S.}~\bibnamefont {Hughes}},\
  }\bibfield  {title} {\enquote {\bibinfo {title} {On‐chip single photon
  sources using planar photonic crystals and single quantum dots},}\
  }\href@noop {} {\bibfield  {journal} {\bibinfo  {journal} {Laser Photonics
  Rev.}\ }\textbf {\bibinfo {volume} {4}},\ \bibinfo {pages} {499--516}
  (\bibinfo {year} {2010})}\BibitemShut {NoStop}%
\bibitem [{\citenamefont {Quilter}\ \emph {et~al.}(2015)\citenamefont
  {Quilter}, \citenamefont {Brash}, \citenamefont {Liu}, \citenamefont
  {Gl{\"a}ssl}, \citenamefont {Barth}, \citenamefont {Axt}, \citenamefont
  {Ramsay}, \citenamefont {Skolnick},\ and\ \citenamefont {Fox}}]{quilter15}%
  \BibitemOpen
  \bibfield  {author} {\bibinfo {author} {\bibfnamefont {J.~H.}\ \bibnamefont
  {Quilter}}, \bibinfo {author} {\bibfnamefont {A.~J.}\ \bibnamefont {Brash}},
  \bibinfo {author} {\bibfnamefont {F.}~\bibnamefont {Liu}}, \bibinfo {author}
  {\bibfnamefont {M.}~\bibnamefont {Gl{\"a}ssl}}, \bibinfo {author}
  {\bibfnamefont {A.~M.}\ \bibnamefont {Barth}}, \bibinfo {author}
  {\bibfnamefont {V.~M.}\ \bibnamefont {Axt}}, \bibinfo {author} {\bibfnamefont
  {A.~J.}\ \bibnamefont {Ramsay}}, \bibinfo {author} {\bibfnamefont {M.~S.}\
  \bibnamefont {Skolnick}}, \ and\ \bibinfo {author} {\bibfnamefont {A.~M.}\
  \bibnamefont {Fox}},\ }\bibfield  {title} {\enquote {\bibinfo {title}
  {{Phonon-Assisted Population Inversion of a Single InGaAs/GaAs Quantum Dot by
  Pulsed Laser Excitation}},}\ }\href@noop {} {\bibfield  {journal} {\bibinfo
  {journal} {Phys. Rev. Lett.}\ }\textbf {\bibinfo {volume} {114}},\ \bibinfo
  {pages} {137401} (\bibinfo {year} {2015})}\BibitemShut {NoStop}%
\bibitem [{\citenamefont {Reindl}\ \emph {et~al.}(2019)\citenamefont {Reindl},
  \citenamefont {Weber}, \citenamefont {Huber}, \citenamefont {Schimpf},
  \citenamefont {Covre~da Silva}, \citenamefont {Portalupi}, \citenamefont
  {Trotta}, \citenamefont {Michler},\ and\ \citenamefont
  {Rastelli}}]{reindl19}%
  \BibitemOpen
  \bibfield  {author} {\bibinfo {author} {\bibfnamefont {Marcus}\ \bibnamefont
  {Reindl}}, \bibinfo {author} {\bibfnamefont {Jonas~H.}\ \bibnamefont
  {Weber}}, \bibinfo {author} {\bibfnamefont {Daniel}\ \bibnamefont {Huber}},
  \bibinfo {author} {\bibfnamefont {Christian}\ \bibnamefont {Schimpf}},
  \bibinfo {author} {\bibfnamefont {Saimon~F.}\ \bibnamefont {Covre~da Silva}},
  \bibinfo {author} {\bibfnamefont {Simone~L.}\ \bibnamefont {Portalupi}},
  \bibinfo {author} {\bibfnamefont {Rinaldo}\ \bibnamefont {Trotta}}, \bibinfo
  {author} {\bibfnamefont {Peter}\ \bibnamefont {Michler}}, \ and\ \bibinfo
  {author} {\bibfnamefont {Armando}\ \bibnamefont {Rastelli}},\ }\bibfield
  {title} {\enquote {\bibinfo {title} {{Highly indistinguishable single photons
  from incoherently and coherently excited GaAs quantum dots}},}\ }\href@noop
  {} {\bibfield  {journal} {\bibinfo  {journal} {arXiv:1901.11251}\ } (\bibinfo
  {year} {2019})}\BibitemShut {NoStop}%
\bibitem [{\citenamefont {Wei}\ \emph {et~al.}(2014)\citenamefont {Wei},
  \citenamefont {He}, \citenamefont {Chen}, \citenamefont {Hu}, \citenamefont
  {He}, \citenamefont {Wu}, \citenamefont {Schneider}, \citenamefont {Kamp},
  \citenamefont {H{\"o}fling}, \citenamefont {Lu},\ and\ \citenamefont
  {Pan}}]{wei14}%
  \BibitemOpen
  \bibfield  {author} {\bibinfo {author} {\bibfnamefont {Y.-J.}\ \bibnamefont
  {Wei}}, \bibinfo {author} {\bibfnamefont {Y.-M.}\ \bibnamefont {He}},
  \bibinfo {author} {\bibfnamefont {M.-C.}\ \bibnamefont {Chen}}, \bibinfo
  {author} {\bibfnamefont {Y.-N.}\ \bibnamefont {Hu}}, \bibinfo {author}
  {\bibfnamefont {Y.}~\bibnamefont {He}}, \bibinfo {author} {\bibfnamefont
  {D.}~\bibnamefont {Wu}}, \bibinfo {author} {\bibfnamefont {C.}~\bibnamefont
  {Schneider}}, \bibinfo {author} {\bibfnamefont {M.}~\bibnamefont {Kamp}},
  \bibinfo {author} {\bibfnamefont {S.}~\bibnamefont {H{\"o}fling}}, \bibinfo
  {author} {\bibfnamefont {C.-Y.}\ \bibnamefont {Lu}}, \ and\ \bibinfo {author}
  {\bibfnamefont {J.-W.}\ \bibnamefont {Pan}},\ }\bibfield  {title} {\enquote
  {\bibinfo {title} {{Deterministic and Robust Generation of Single Photons
  from a Single Quantum Dot with 99.5\% Indistinguishability Using Adiabatic
  Rapid Passage}},}\ }\href@noop {} {\bibfield  {journal} {\bibinfo  {journal}
  {Nano Lett.}\ }\textbf {\bibinfo {volume} {14}},\ \bibinfo {pages} {6515}
  (\bibinfo {year} {2014})}\BibitemShut {NoStop}%
\bibitem [{\citenamefont {Reuble}\ \emph {et~al.}(2014)\citenamefont {Reuble},
  \citenamefont {Dilcher}, \citenamefont {Gamouras}, \citenamefont
  {Ramachandran}, \citenamefont {Yi~Shi~Yang}, \citenamefont {Freisem},
  \citenamefont {Deppe},\ and\ \citenamefont {Hall}}]{reuble14}%
  \BibitemOpen
  \bibfield  {author} {\bibinfo {author} {\bibfnamefont {Mathew}\ \bibnamefont
  {Reuble}}, \bibinfo {author} {\bibfnamefont {Eric}\ \bibnamefont {Dilcher}},
  \bibinfo {author} {\bibfnamefont {Angela}\ \bibnamefont {Gamouras}}, \bibinfo
  {author} {\bibfnamefont {Ajan}\ \bibnamefont {Ramachandran}}, \bibinfo
  {author} {\bibfnamefont {Hong}\ \bibnamefont {Yi~Shi~Yang}}, \bibinfo
  {author} {\bibfnamefont {Sabine}\ \bibnamefont {Freisem}}, \bibinfo {author}
  {\bibfnamefont {Dennis}\ \bibnamefont {Deppe}}, \ and\ \bibinfo {author}
  {\bibfnamefont {Kimberly~C.}\ \bibnamefont {Hall}},\ }\bibfield  {title}
  {\enquote {\bibinfo {title} {{Subpicosecond adiabatic rapid passage on a
  single semiconductor quantum dot: Phonon-mediated dephasing in the
  strong-driving regime}},}\ }\href@noop {} {\bibfield  {journal} {\bibinfo
  {journal} {Phys. Rev. B}\ }\textbf {\bibinfo {volume} {90}},\ \bibinfo
  {pages} {035316} (\bibinfo {year} {2014})}\BibitemShut {NoStop}%
\bibitem [{\citenamefont {Reiter}\ \emph {et~al.}(2014)\citenamefont {Reiter},
  \citenamefont {Kuhn}, \citenamefont {Gl\"{a}ssl},\ and\ \citenamefont
  {Axt}}]{reiter14}%
  \BibitemOpen
  \bibfield  {author} {\bibinfo {author} {\bibfnamefont {D.~E.}\ \bibnamefont
  {Reiter}}, \bibinfo {author} {\bibfnamefont {T.}~\bibnamefont {Kuhn}},
  \bibinfo {author} {\bibfnamefont {M}~\bibnamefont {Gl\"{a}ssl}}, \ and\
  \bibinfo {author} {\bibfnamefont {V.~M.}\ \bibnamefont {Axt}},\ }\bibfield
  {title} {\enquote {\bibinfo {title} {The role of phonons for exciton and
  biexciton generation in an optically driven quantum dot},}\ }\href@noop {}
  {\bibfield  {journal} {\bibinfo  {journal} {J. Phys. Condens. Matter}\
  }\textbf {\bibinfo {volume} {26}},\ \bibinfo {pages} {423203} (\bibinfo
  {year} {2014})}\BibitemShut {NoStop}%
\bibitem [{\citenamefont {Roy}\ and\ \citenamefont {Hughes}(2011)}]{roy11}%
  \BibitemOpen
  \bibfield  {author} {\bibinfo {author} {\bibfnamefont {C.}~\bibnamefont
  {Roy}}\ and\ \bibinfo {author} {\bibfnamefont {S.}~\bibnamefont {Hughes}},\
  }\bibfield  {title} {\enquote {\bibinfo {title} {{Phonon-Dressed Mollow
  Triplet in the Regime of Cavity Quantum Electrodynamics: Excitation-Induced
  Dephasing and Nonperturbative Cavity Feeding Effects}},}\ }\href@noop {}
  {\bibfield  {journal} {\bibinfo  {journal} {Phys. Rev. Lett.}\ }\textbf
  {\bibinfo {volume} {106}},\ \bibinfo {pages} {247403} (\bibinfo {year}
  {2011})}\BibitemShut {NoStop}%
\bibitem [{\citenamefont {Roy}\ and\ \citenamefont {Hughes}(2012)}]{roy12}%
  \BibitemOpen
  \bibfield  {author} {\bibinfo {author} {\bibfnamefont {C.}~\bibnamefont
  {Roy}}\ and\ \bibinfo {author} {\bibfnamefont {S.}~\bibnamefont {Hughes}},\
  }\bibfield  {title} {\enquote {\bibinfo {title} {{Polaron master equation
  theory of the quantum-dot Mollow triplet in a semiconductor cavity-QED
  system}},}\ }\href@noop {} {\bibfield  {journal} {\bibinfo  {journal} {Phys.
  Rev. B}\ }\textbf {\bibinfo {volume} {85}},\ \bibinfo {pages} {115309}
  (\bibinfo {year} {2012})}\BibitemShut {NoStop}%
\bibitem [{\citenamefont {McCutcheon}\ \emph {et~al.}(2011)\citenamefont
  {McCutcheon}, \citenamefont {Dattani}, \citenamefont {Gauger}, \citenamefont
  {Lovett},\ and\ \citenamefont {Nazir}}]{mccutcheon11}%
  \BibitemOpen
  \bibfield  {author} {\bibinfo {author} {\bibfnamefont {D.~P.~S.}\
  \bibnamefont {McCutcheon}}, \bibinfo {author} {\bibfnamefont {N.~S.}\
  \bibnamefont {Dattani}}, \bibinfo {author} {\bibfnamefont {E.~M.}\
  \bibnamefont {Gauger}}, \bibinfo {author} {\bibfnamefont {B.~W.}\
  \bibnamefont {Lovett}}, \ and\ \bibinfo {author} {\bibfnamefont
  {A.}~\bibnamefont {Nazir}},\ }\bibfield  {title} {\enquote {\bibinfo {title}
  {{A general approach to quantum dynamics using a variational master equation:
  Application to phonon-damped Rabi rotations in quantum dots}},}\ }\href@noop
  {} {\bibfield  {journal} {\bibinfo  {journal} {Phys. Rev. B}\ }\textbf
  {\bibinfo {volume} {84}},\ \bibinfo {pages} {081305(R)} (\bibinfo {year}
  {2011})}\BibitemShut {NoStop}%
\bibitem [{\citenamefont {McCutcheon}\ and\ \citenamefont
  {Nazir}(2010)}]{mccutcheon10}%
  \BibitemOpen
  \bibfield  {author} {\bibinfo {author} {\bibfnamefont {D.~P.~S.}\
  \bibnamefont {McCutcheon}}\ and\ \bibinfo {author} {\bibfnamefont
  {A.}~\bibnamefont {Nazir}},\ }\bibfield  {title} {\enquote {\bibinfo {title}
  {{Quantum dot Rabi rotations beyond the weak exciton-phonon coupling
  regime}},}\ }\href@noop {} {\bibfield  {journal} {\bibinfo  {journal} {New J.
  Phys.}\ }\textbf {\bibinfo {volume} {12}},\ \bibinfo {pages} {113042}
  (\bibinfo {year} {2010})}\BibitemShut {NoStop}%
\bibitem [{\citenamefont {Trivedi}\ \emph {et~al.}(2019)\citenamefont
  {Trivedi}, \citenamefont {Fischer}, \citenamefont {Vuckovic},\ and\
  \citenamefont {Muller}}]{trivedi19}%
  \BibitemOpen
  \bibfield  {author} {\bibinfo {author} {\bibfnamefont {Rahul}\ \bibnamefont
  {Trivedi}}, \bibinfo {author} {\bibfnamefont {Kevin}\ \bibnamefont
  {Fischer}}, \bibinfo {author} {\bibfnamefont {Jelena}\ \bibnamefont
  {Vuckovic}}, \ and\ \bibinfo {author} {\bibfnamefont {Kai}\ \bibnamefont
  {Muller}},\ }\bibfield  {title} {\enquote {\bibinfo {title} {Generation of
  non-classical light using semiconductor quantum dots},}\ }\href@noop {}
  {\bibfield  {journal} {\bibinfo  {journal} {arXiv:1901.06367}\ } (\bibinfo
  {year} {2019})}\BibitemShut {NoStop}%
\bibitem [{\citenamefont {Nazir}\ and\ \citenamefont
  {McCutcheon}(2016)}]{nazir16}%
  \BibitemOpen
  \bibfield  {author} {\bibinfo {author} {\bibfnamefont {A.}~\bibnamefont
  {Nazir}}\ and\ \bibinfo {author} {\bibfnamefont {D.~P.~S.}\ \bibnamefont
  {McCutcheon}},\ }\bibfield  {title} {\enquote {\bibinfo {title} {Modelling
  exciton-phonon interactions in optically driven quantum dots},}\ }\href@noop
  {} {\bibfield  {journal} {\bibinfo  {journal} {J. Phys. Condens. Matter}\
  }\textbf {\bibinfo {volume} {28}},\ \bibinfo {pages} {103002} (\bibinfo
  {year} {2016})}\BibitemShut {NoStop}%
\bibitem [{\citenamefont {Hargart}\ \emph {et~al.}(2016)\citenamefont
  {Hargart}, \citenamefont {M{\"u}ller}, \citenamefont {Roy-Choudhury},
  \citenamefont {Portalupi}, \citenamefont {Schneider}, \citenamefont
  {H{\"o}fling}, \citenamefont {Kamp}, \citenamefont {Hughes},\ and\
  \citenamefont {Michler}}]{hargart16}%
  \BibitemOpen
  \bibfield  {author} {\bibinfo {author} {\bibfnamefont {F.}~\bibnamefont
  {Hargart}}, \bibinfo {author} {\bibfnamefont {M.}~\bibnamefont {M{\"u}ller}},
  \bibinfo {author} {\bibfnamefont {K.}~\bibnamefont {Roy-Choudhury}}, \bibinfo
  {author} {\bibfnamefont {S.~L.}\ \bibnamefont {Portalupi}}, \bibinfo {author}
  {\bibfnamefont {C.}~\bibnamefont {Schneider}}, \bibinfo {author}
  {\bibfnamefont {S.}~\bibnamefont {H{\"o}fling}}, \bibinfo {author}
  {\bibfnamefont {M.}~\bibnamefont {Kamp}}, \bibinfo {author} {\bibfnamefont
  {S.}~\bibnamefont {Hughes}}, \ and\ \bibinfo {author} {\bibfnamefont
  {P.}~\bibnamefont {Michler}},\ }\bibfield  {title} {\enquote {\bibinfo
  {title} {Cavity enhanced simultaneous dressing of quantum dot exciton and
  biexciton states},}\ }\href@noop {} {\bibfield  {journal} {\bibinfo
  {journal} {Phys. Rev. B}\ }\textbf {\bibinfo {volume} {93}},\ \bibinfo
  {pages} {115308} (\bibinfo {year} {2016})}\BibitemShut {NoStop}%
\bibitem [{\citenamefont {Hughes}\ \emph {et~al.}(2019)\citenamefont {Hughes},
  \citenamefont {Franke}, \citenamefont {Gustin}, \citenamefont {Dezfouli},
  \citenamefont {Knorr},\ and\ \citenamefont {Richter}}]{hughes19}%
  \BibitemOpen
  \bibfield  {author} {\bibinfo {author} {\bibfnamefont {Stephen}\ \bibnamefont
  {Hughes}}, \bibinfo {author} {\bibfnamefont {Sebastian}\ \bibnamefont
  {Franke}}, \bibinfo {author} {\bibfnamefont {Chris}\ \bibnamefont {Gustin}},
  \bibinfo {author} {\bibfnamefont {Mohsen~Kamandar}\ \bibnamefont {Dezfouli}},
  \bibinfo {author} {\bibfnamefont {Andreas}\ \bibnamefont {Knorr}}, \ and\
  \bibinfo {author} {\bibfnamefont {Marten}\ \bibnamefont {Richter}},\
  }\bibfield  {title} {\enquote {\bibinfo {title} {Theory and limits of
  on-demand single photon sources using plasmonic resonators: a quantized
  quasinormal mode approach},}\ }\href@noop {} {\bibfield  {journal} {\bibinfo
  {journal} {ACS Photonics [in press]}\ } (\bibinfo {year} {2019})}\BibitemShut
  {NoStop}%
\bibitem [{\citenamefont {Carmichael}(1999)}]{carmichael}%
  \BibitemOpen
  \bibfield  {author} {\bibinfo {author} {\bibfnamefont {H.~J.}\ \bibnamefont
  {Carmichael}},\ }\href@noop {} {\emph {\bibinfo {title} {Statistical Methods
  in Quantum Optics 1: Master Equations and Fokker-Planck Equations}}}\
  (\bibinfo  {publisher} {Springer-Verlag, Berlin},\ \bibinfo {year}
  {1999})\BibitemShut {NoStop}%
\bibitem [{\citenamefont {Roy-Choudhury}\ and\ \citenamefont
  {Hughes}(2015)}]{roychoudhury15}%
  \BibitemOpen
  \bibfield  {author} {\bibinfo {author} {\bibfnamefont {K.}~\bibnamefont
  {Roy-Choudhury}}\ and\ \bibinfo {author} {\bibfnamefont {S.}~\bibnamefont
  {Hughes}},\ }\bibfield  {title} {\enquote {\bibinfo {title} {{Spontaneous
  emission from a quantum dot in a structured photonic reservoir:
  Phonon-mediated breakdown of Fermi’s golden rule}},}\ }\href@noop {}
  {\bibfield  {journal} {\bibinfo  {journal} {Optica}\ }\textbf {\bibinfo
  {volume} {2}},\ \bibinfo {pages} {434} (\bibinfo {year} {2015})}\BibitemShut
  {NoStop}%
\bibitem [{\citenamefont {Cosacchi}\ \emph {et~al.}(2019)\citenamefont
  {Cosacchi}, \citenamefont {Ungar}, \citenamefont {Cygorek}, \citenamefont
  {Vagov},\ and\ \citenamefont {Axt}}]{cosacchi19}%
  \BibitemOpen
  \bibfield  {author} {\bibinfo {author} {\bibfnamefont {M.}~\bibnamefont
  {Cosacchi}}, \bibinfo {author} {\bibfnamefont {F.}~\bibnamefont {Ungar}},
  \bibinfo {author} {\bibfnamefont {M}~\bibnamefont {Cygorek}}, \bibinfo
  {author} {\bibfnamefont {A.}~\bibnamefont {Vagov}}, \ and\ \bibinfo {author}
  {\bibfnamefont {V.~M.}\ \bibnamefont {Axt}},\ }\bibfield  {title} {\enquote
  {\bibinfo {title} {{Emission-Frequency Separated High Quality Single-Photon
  Sources Enabled by Phonons}},}\ }\href@noop {} {\bibfield  {journal}
  {\bibinfo  {journal} {Phys. Rev. Lett.}\ }\textbf {\bibinfo {volume} {123}},\
  \bibinfo {pages} {017403} (\bibinfo {year} {2019})}\BibitemShut {NoStop}%
\end{thebibliography}%
        \end{document}